\documentclass[a4paper,11pt]{article}
\usepackage{jheppub} 
\usepackage[normalem]{ulem}
\usepackage{slashed}
\usepackage{bigints}

\def\be{\begin{equation}}
\def\ee{\end{equation}}
\def\figs/B{B}
\def\bea{\begin{eqnarray}}
\def\eea{\end{eqnarray}}
\def\bg{\begin{eqnarray}}
\def\nd{\end{eqnarray}}

\def\log{{\rm log}}
\def\ln{{\rm log}}

\newcommand{\inter}{{\mbox{\tiny int}}}
\newcommand{\kr}{{\mbox{\tiny KR}}}
\newcommand{\dph}{{\mbox{\tiny DP}}}

\newcommand{\pl}{{\mbox{\tiny P}}}
\newcommand{\eff}{{\mbox{\tiny eff}}}

\newcommand{\rh}{{\mbox{\tiny RH}}}

\title{\boldmath  
Cosmological Implications of Kalb-Ramond-Like Particles}
\author[a]{Christian Capanelli,}
\author[b]{Leah Jenks,}
\author[b,c]{Edward W. Kolb,}
\author[d]{Evan McDonough}

\affiliation[a]{Department of Physics \& Trottier Space Institute,\\ 
McGill University, Montr\'eal, QC H3A 2T8, Canada.}
\affiliation[b]{Kavli Institute for Cosmological Physics, \\
The University of Chicago,  5640 South Ellis Avenue, Chicago, IL 60637, U.S.A.}
\affiliation[c]{Enrico Fermi Institute,\\
The University of Chicago,  5640 South Ellis Avenue, Chicago, IL 60637, U.S.A.}
\affiliation[d]{Department of Physics,\\ University of Winnipeg, Winnipeg MB, R3B 2E9, Canada}

\emailAdd{christian.capanelli@mail.mcgill.ca}
\emailAdd{ljenks@uchicago.edu}
\emailAdd{Rocky.Kolb@uchicago.edu}
\emailAdd{e.mcdonough@uwinnipeg.ca}

\abstract{
The Kalb-Ramond field is an antisymmetric, rank-two tensor field which most notably appears in the context of string theory, but has largely been unexplored in the context of cosmology. In this work, motivated by the Kalb-Ramond field in string theory, and antisymmetric tensor fields that emerge in effective field theories ranging from particle physics to condensed matter,
we study the primordial production of interacting massive Kalb-Ramond-like-particles (KRLPs). 
 KRLPs contain features of both dark photon and axion models, which can be appreciated via their duality properties. While the \emph{massless} non-interacting KRLP is dual to a pseudoscalar, and the \emph{massive} non-interacting KRLP is dual to a pseudovector, the interacting massive KRLP can be distinguished from its scalar and vector counterparts. We study early-universe production of KRLPs via the freeze-in mechanism, considering a `dark photon-like' interaction, an `axion-like' interaction, and a `Higgs portal' interaction, as well as production via cosmological gravitational particle production. We find that as a dark matter candidate, KRLPs  can be produced by all of the above mechanisms and account for the relic density of dark matter today for a wide range of masses. Finally, we comment on the potential to obtain both warm and cold dark matter subcomponents, and speculate on observational and experimental prospects. 
}


\begin{document}
 \maketitle
\flushbottom

\section{Introduction}
\label{sec:intro}

The Kalb-Ramond (KR) field, $B_{\mu\nu}$, is an antisymmetric, rank-2 tensor field. It was first introduced in 1974 within the context of string theory by  Kalb and Ramond to describe the interactions between fundamental strings \cite{Kalb:1974yc}, but it also arises generically as a $(1,0)\oplus(0,1)$ representation of the Lorentz group in four dimensions.\footnote{Note that the antisymmetric two-form field was first introduced as the `notoph' in \cite{Ogievetsky:1966eiu}. See e.g., \cite{Ivanov:2016lha} for a discussion.} The KR field can be thought of as a generalization of the electromagnetic potential, with one key difference: While the electromagnetic potential couples to point particles, the KR field couples to extended objects, e.g., strings. Historically, the KR field has been closely associated with the axion, sometimes being referred to as the `Kalb-Ramond axion,' due to a duality which maps directly from the massless non-interacting KR field to a pseuduscalar field \cite{Svrcek:2006yi}. Additionally, it has been studied in the context of Lorentz violating gauge theories \cite{Kostelecky:1989jp, Higashijima:2001sq, Altschul:2009ae,Hernaski:2016dyk,Assuncao:2019azw,Almeida:2020lsn}. The massive KR field on the other hand, has been less well studied but also possesses a duality: the non-interacting massive KR is dual to a massive pseudovector field \cite{Hell:2021wzm} (see \cite{Smailagic:2001ch} for related earlier work). In this work, we focus on the massive interacting Kalb-Ramond field of string theory as well as a larger class of antisymmetric rank-2 tensors arising in a variety of effective field theories, which we refer to as \textit{Kalb-Ramond-Like Particles} (KRLPs). 

Despite the marked presence of the KR field and KRLPs in string theory, as well as as a representation of the Lorentz group, little work has been done to study either more broadly in the context of cosmology. As mentioned above, in its original form, the KR couples to the fundamental strings of string theory and has been well studied in the context of string theory. However, it has also been suggested that that the KR can couple to cosmic strings \cite{Vilenkin:1986ku} and have implications for the early universe \cite{Maity:2004he,BeltranAlmeida:2018nin,Almeida:2019xzt,Mavromatos:2020kzj,Mavromatos:2021urx,Basilakos:2019acj} and leptogenesis with a background KR field \cite{Mavromatos:2023bdx,Mavromatos:2020dkh,Mavromatos:2012ii,Bossingham:2017gtm,deCesare:2014dga}. Furthermore, previous work has considered the KR field coupled to dark matter as a portal to the standard model (SM) \cite{Dashko:2018dsw, Dick:2020eop}, in the context of bouncing cosmology \cite{DeRisi:2007dn, Nair:2021hnm, Paul:2022mup} dark energy \cite{BeltranAlmeida:2019fou,Orjuela-Quintana:2022jrg} and in the gravitational sector \cite{Lessa:2019bgi, Kumar:2020hgm,Chakraborty:2016lxo}. Building upon the above efforts, our goal for this work is to consider implications of KRLPs in the context of cosmology. As a particular example, we study KRLPs as dark matter (DM) candidates. 

Dark matter makes up nearly a quarter of the energy density of the universe, while its exact nature is still unknown. There exist myriad theoretical proposals; however, experimental efforts to confirm any one of them have thus far led to null results. In this work, we consider the KRLP as a dark matter candidate which shares many properties of two established dark matter candidates: the axion and the dark photon. The axion is a light pseudo-scalar field which was originally introduced as a solution to the strong CP problem in quantum chromodynamics (QCD) \cite{Peccei:1977hh, Peccei:1977ur, Wilczek:1977pj, Weinberg:1977ma}, but has since been generalized to a larger class of axion-like-particles (ALPs) which have gained significant attention as a DM candidate. Axions are particularly attractive as a DM candidate due to the fact that a spectrum, or `axiverse,' of particles is predicted by string theory \cite{Svrcek:2006yi, Arvanitaki:2009fg,Cicoli:2012sz} and by more conventional field theories, such as dark QCD, where an axiverse emerges from the spectrum of dark pions \cite{Alexander:2023wgk,Maleknejad:2022gyf}. Furthermore, dark matter axions inhabit a wide potential mass range, from $\mathcal{O}$(MeV) down to the ultralight regime of $\mathcal{O}(10^{-22})$ eV \cite{Hui:2021tkt, Marsh:2015xka} (or lower, for a subcomponent of the DM) and can be produced by several mechanisms including via freeze-in \cite{Langhoff:2022bij}. The dark photon, on the other hand, is a massive vector field which has been suggested as both a dark matter candidate itself, as well as a portal connecting the SM to other dark matter particles \cite{Redondo_2009, Nelson:2011sf,Long:2019lwl,Arias:2012az,Graham:2015rva, Kolb:2020fwh,Co:2018lka,Dror:2018pdh,Bastero-Gil:2018uel, Bastero-Gil:2021wsf, Agrawal:2018vin, Krnjaic:2022wor}. It can arise due to kinetic mixing with the SM photon, and spans a wide parameter space in mass and coupling \cite{Pospelov:2008jk, Fradette_2014,Jaeckel_2013}.

Our study of KRLP DM includes aspects of both axion and dark photon DM and is complementary to these existing models. We first emphasize the close analogy between the axion and the `axiverse' arising in string theory to the origins of the KR field and KRLPs (though as for the axion, the existence of KRLPs is not solely tied to string theory). The massive KRLP can have axial couplings to the standard model (SM) particles. However, these interactions are distinct from the axion, because the KRLP is a spin-1 particle, not a scalar. Similarly, the duality to a spin-1 field gives the KRLP properties similar to the dark photon, but which can be distinguished due to the fact that the KRLP duality is to a \emph{pseudovector}, while the dark photon is a \emph{vector}. Another distinction from the dark photon, as we will see, lies in the structure of the allowed couplings to Standard Model, as a simple consequence of Lorentz invariance and the nature of the KR field as two-index object vs.\ the one Lorentz index of the dark photon.

In this work, we study several production mechanisms for KRLP dark matter. We primarily focus on KRLP production via freeze-in \cite{Hall:2009bx}, in which the KRLP is coupled to SM fermions. We consider two sets of interactions coupling to SM fermions, which we denote as `dark photon-like,' and `axion-like.'  These two classes of interaction are defined as a direct coupling of the gauge field ($A_\mu$ or $B_{\mu \nu}$) vs.\ a derivative coupling to the SM axial vector current for dark photon-like and axion-like interactions respectively. 

For the axion-like interaction, we find that the freeze-in of KRLPs leads to a viable parameter space for the coupling and mass which is qualitatively the same as for freeze-in of axion-like particles. On the other hand, freeze-in of KRLPs via the dark photon-like interaction qualitatively differs from the dark photon case. This distinction is due to a coupling to the non-conservation of the SM fermion tensor current 
in the former case, versus the conserved vector current in the latter. For the dark photon-like interaction, the KRLPs have enhanced production in the low-mass regime leading to a large region of the coupling-mass parameter space in which KRLPs are overproduced. Furthermore, freeze-in from both of these interactions can also lead to a warm dark matter subcomponent, analogous to \cite{Dvorkin:2020xga}. 

Beyond freeze-in via coupling to SM fermions, we also consider KRLP production via a Higgs portal freeze-in scenario as well as gravitational particle production (GPP). Both cases are characterized by quadratic couplings in the dark photon or KR field (i.e., coupling of $A_{\mu}A^\mu$ or $B_{\mu \nu} B^{\mu \nu}$), making them insensitive to the parity properties which distinguish the two fields. Indeed, in both of these cases, we find the production of KRLPs is indistinguishable from standard vector dark matter. Thus, the viable parameter space to account for all of the DM via these production mechanisms is the same as for the dark photon, and dark photon results from studies such as \cite{Kolb:2017jvz, Kolb:2020fwh, s1-NM} can also be applied to KRLPs.

Our results imply that KRLPs are viable dark matter candidates which combine attractive features of both dark photon and axion models. In particular, due to the close relationship between the models, our results indicate current experimental efforts searching for both the axion and the dark photon may also be sensitive to KRLPs.

 The structure of the paper is as follows. We begin with a discussion of  the free KR field in Sec.~\ref{sec:background}, where we discuss the duality of the free KR field to axions and pseudovectors, as well as the differences between the KR and Proca fields, particularly in taking the massless limit of the massive KR vs. Proca field. Further details on the quantization and polarization states of KR are given in App.~\ref{app:KR}. In Sec.~\ref{sec:Interactions} we enumerate the interactions that the KR field may have with the standard model and other particles, arising from both string theory and more general effective field theory contexts. For the reader interested in further technical details, we discuss the string theory origins of the KR field in Sec.~\ref{Sec:Strings}, and propose mechanisms for generating a massive KR or KRLP in string theory. We then turn to a study of the production of KRLPs as a DM candidate.  Our main results for KRLP dark matter produced via freeze-in are presented in Sec.~\ref{sec:freezein}. We also comment on gravitational particle production as an additional production mechanism in Sec.~\ref{sec:other}. Finally, we discuss `frequently asked questions' about observational and experimental prospects in Sec.~\ref{sec:FAQ}, and conclude with a discussion and directions for future work in Sec.~\ref{sec:Conclusions}. For the remainder of the paper, the following conventions are used: We assume a $(+, -, -, -)$ metric signature and work in natural units, $c = \hbar = 1$.

\section{The Kalb-Ramond Field}

\label{sec:background}

\subsection{Kalb-Ramond as the Union of Axion and Proca Fields}

\label{sec:backgroundKR}
  As we have discussed, the Kalb-Ramond arises from string theory, however we can also consider KRLPs as general antisymmetric tensor fields in the context of the Lorentz group. In particular,  $B_{\mu \nu}$ corresponds to the rank-2 $(0,1)\oplus (1,0)$ tensor representation of the Lorentz group. This can be understood from decomposing any rank-two tensor, $T_{\mu\nu}$, into a trace piece, antisymmetric piece, and a symmetric and traceless piece. Of these, the antisymmetric piece, denoted by $B_{\mu\nu}$, is the antisymmetric tensor representation of the Lorentz group. 
 Thus, in addition to motivation of the KR field from string theory, one can understand KRLPs as a well defined field theory in four dimensions, regardless of their stringy origin .

The KR field of string theory was first proposed by Kalb and Ramond in 1974 to describe the interactions of strings \cite{Kalb:1974yc}, but the following discussion generically holds for any antisymmetric, rank-two tensor, or KRLP. The free action for the massive KRLP is given by 
\be 
    S_\kr = \frac{1}{12}\int d^4x \left(H_{\mu\nu\rho}H^{\mu\nu\rho} - 3m^2 B_{\mu\nu}B^{\mu\nu}\right), 
\label{eq:SKR}
\ee 
where $B_{\mu\nu}$ is the antisymmetric KRLP field, $m$ is the mass, and $H_{\mu \nu \rho}$ is the KRLP field strength, given by
\be 
H_{\mu\nu\rho} = \partial_{\mu}B_{\nu\rho}+\partial_{\nu}B_{\rho\mu}+\partial_{\rho}B_{\mu\nu}.
\ee 
In the massless limit, the Kalb-Ramond action admits a local gauge invariance,
\begin{equation}
    B_{\mu \nu}\rightarrow B_{\mu \nu} + F_{\mu \nu}
\end{equation}
where $F_{\mu \nu}= \partial_\mu V_{\nu} - \partial_\nu V_{\mu}$ for an arbitrary four-vector $V_{\mu}(x)$. Consider the symmetries of the KRLP under parity. Unlike a vector, which is odd under parity, The KRLP, $B_{\mu \nu}$ is parity-even. This contrasts with a conventional gauge field $A_{\mu}$, which, as a conventional vector, is odd under parity. This implies that the KRLP field strength $H_{\mu \nu \sigma}$ is parity odd, consistent with the identification of $H_{\mu \nu \sigma}$ as a source of ``torsion'' \cite{Strominger:1986uh,Becker:2006dvp},\footnote{C.f. Eq.~10.206 of \cite{Becker:2006dvp}.} and, in the massless limit, the identification of the Kalb-Ramond ``axion'' $\theta$, defined by $\partial_{\mu}\theta =\epsilon_{\mu \nu \sigma \rho} H^{\nu \sigma \rho} $  , as a pseudoscalar. Finally, the dual field strength $(\star H)^{\mu}$ is even under parity, making it a pseudo-vector.

The equations of motion for the massive KRLP are given by 
\be 
\label{eq:KREOM}
\partial_\mu H^{\mu\nu\rho} = m^2B^{\nu\rho} ,
\ee 
analogous to that of a massive vector (Proca) field,
\begin{equation}
    \partial_{\mu}F^{\mu \nu} = m^2 A^{\nu}.
\end{equation}
A textbook discussion of the latter can be found in \cite{Schwartz:2014sze}. Taking the derivative of the Proca equation of motion leads to a constraint equation,
\begin{equation}
    m^2 \partial_\mu A^{\mu}=0
\end{equation}
which removes one propagating degree of freedom, leaving the 3 degrees of freedom of a massive vector field. This also serves as the defining relation for polarization vectors,
\begin{equation}
    \partial_\mu \varepsilon^{(\lambda)\mu} =0 ,
\end{equation}
with $\lambda=1,2,3$ the three polarization states,  which together form a basis for solutions to the Proca equation of motion. 

We can straightforwardly apply this analysis to the KRLP. Taking the derivative of the equation of motion, Eq.~\eqref{eq:KREOM} yields a constraint equation,
\begin{equation}
\label{eq:2.8}
    m^2 \partial_\mu B^{\mu \nu}=0,
\end{equation}
which serves as the defining relation of the KR polarization {\it tensors} $\varepsilon^{\mu \nu}$. We explicitly construct the KR polarization tensors, and the corresponding sum over polarization states, in App.~\ref{app:KR}.

The constraint Eq.~\eqref{eq:2.8} removes 3 on-shell degrees of freedom\footnote{Note that while Eq.~\eqref{eq:2.8} is 4 equations, only 3 of these are linearly independent. } from the of 6 a priori independent components of $B^{\mu \nu}$, leaving 3 remaining on-shell degrees of freedom. This suggests a duality (see Ref.~\cite{Hell:2021wzm}) to a massive vector field, which also propagates 3 degrees of freedom. A key distinction however, is that the massless limit of a conventional massive vector field (a Proca field) is a massless vector, whereas the massless KRLP is dual to a pseudo-scalar: the famous the Kalb-Ramond axion of string theory \cite{Svrcek:2006yi}. This suggests that the KRLP shares features of both the spin-one pseudo-vector and axion fields. In what follows, and throughout this work, we will explore this in detail.

\subsection{Duality}

As noted above, the massless KR field is well known in the context of its duality to an axion-like particle,\footnote{Note that while formally the KR is related to an axion-like particle (ALP), not the QCD axion, keeping with convention in the literature, we will refer to these ALPs as simply `axions' in what follows.} namely the `Kalb-Ramond Axion' \cite{Svrcek:2006yi}. Specifically, it is the \textit{massless} KR field which is dual to the axion. One can see this as follows: recall that the equation of motion for the massless  KRLP is given by,
\begin{equation}
    \partial_\mu H^{\mu \nu \rho}=0.
\end{equation}
The general solution to this is,
\begin{equation}
\label{eq:solH}
H^{\mu \nu \rho} =\epsilon^{\mu \nu \rho \sigma}\partial_\sigma \theta
\end{equation}
for a pseudoscalar $\theta$. Inserting this into the free action, one arrives at the action for a free scalar field,
\begin{equation}
     S = \int {\rm d}^{4}x \sqrt{-g} \, \frac{1}{2}(\partial _\mu \theta)^2 .
\end{equation}
The field $\theta$ is the Kalb-Ramond axion, sometimes referred to as the `model-independent' axion of string theory \cite{Svrcek:2006yi}.  Eq.~\eqref{eq:solH} may alternatively be be written as a defining relation for $\theta$,
\begin{equation}
    \partial_\mu \theta \equiv  (\star H)_\mu
\end{equation}
where $\star$ denotes the Hodge dual. Thus the massless KRLP field strength is literally `dual' to a pseudoscalar field, namely the Kalb-Ramond axion. Notice that for a massive KRLP, this duality to an axion fails. The equation of motion, Eq.~\eqref{eq:KREOM}, is no longer solved by $\partial \theta  = \star H$. Inserting this into the equation of motion, one finds,
\begin{equation}
    \epsilon^{\mu \nu \rho \sigma}\partial_\mu \partial_\nu \theta = m^2 B^{\rho \sigma}.
\end{equation}
The left hand side vanishes due to antisymmetry in $\mu$ and $\nu$, indicating this ansatz cannot solve the equation of motion. 

Instead, one finds that the massive KRLP is dual to a massive pseudo-vector field or ``axial vector.''\footnote{Conversely, a massive axion is dual not to a Kalb-Ramond, but is instead dual to a massive three-form gauge field  \cite{Dvali:2005an, Sakhelashvili:2021eid}.} This duality can be seen explicitly via the decomposition of $B_{\mu\nu}$ into temporal and spatial components. Following the discussion in \cite{Hell:2021wzm}, we decompose the spatial components $B_{ij}$ into a pseudovector $B_k$,
\begin{equation}
     B_{ij} = \epsilon_{ijk}B_k, \label{eq:Bk}
\end{equation}
where we note that $\epsilon_{ijk}$ is parity-even (Recall that per its identity as a pseudotensor, it has opposite parity to a rank-3 tensor), as is $B_{ij}$, making $B_k$ necessarily parity even -- a pseudo-vector. The pseudo-vector $B_k$ and the remaining nonzero components of $B_{\mu\nu}$ can be further decomposed as
\begin{align}
    B_i &= B_i^T + \partial_i \phi,\\
    B_{0i} &= C_i^T + \partial_i\mu.
\end{align}
The fields $C_i^T$ and $B_i^T$ satisfy $\partial_i B_i^T  = \partial_i C_i^T  = 0$. Expanding out the action with the above decomposition, one finds that $C_i^T$ and $\mu$ are non-propagating degrees of freedom and can thus be integrated out of the action. Upon doing so and canonically normalizing $B_i^T$ and $\phi$, one can identify $B^T$ and $\phi$ with the propagating transverse and longitudinal modes, respectively. With this decomposition, the action Eq. \eqref{eq:SKR} exactly reduces to the Proca action for a conventional vector field $B_k$ \cite{Hell:2021wzm}.

However, with the above discussion, we emphasize that the KR field is a distinct physical object from the Proca field. Recall that while the KRLP is dual to a parity-even pseudo-vector, which is indistinguishable from a parity-odd vector in the Proca action, in general this may not be the case. Furthermore, one can appreciate that the duality relation, Eq.~\eqref{eq:Bk}, is not Lorentz invariant. As a consequence, the Lorentz-invariant interacting theories have different building blocks: a Lorentz invariant interacting action for KRLPs may be built out of $H^{\mu \nu \sigma}$, $(\star H)^\mu$, and $B^{\mu \nu}$, whereas the Lorentz invariant action for an interacting Proca field may be built out of $F_{\mu \nu}$, $(\star F)^{\mu \nu}$, or $A_\mu$, and there is no Lorentz invariant duality between these various objects. Lastly, the delineation between KRLPs and Proca can be further understood from the massless limit, wherein one finds that the KRLP becomes a massless vector and a {\it pseudo}scalar, in contrast with Proca, which decomposes into a massless vector and a (parity-even) scalar. We discuss details of the massless limit for both the KRLP and Proca cases below.

\subsection{Massless Limit}

To better understand the distinction between the Proca theory and massive KRLPs, we consider the massless limit of each.

\subsubsection{Massless Limit of Massive Vector Field}

The massless limit of the Proca field is well understood. The three degrees of freedom of the Proca field correspond to a scalar field and a massless vector field. The massless theory ostensibly propagates two degrees of freedom, suggesting a discontinuity (the vDVZ discontinuity) in the number of degrees of freedom, however a more careful accounting, e.g. by the `Stueckelberg trick' \cite{Hinterbichler:2011tt}, one may appreciate that the limit is smooth, with apparent reduction of degrees of freedom corresponding simply to a decoupling of the scalar.

Let us see the Stueckelberg trick in detail in the case of the Proca theory. The action can be expressed as,
\begin{equation}\label{massivevect}
S=\int d^4x\left( -\frac{1}{4} F_{\mu\nu}F^{\mu\nu}-\frac{1}{2}m^2 A_\mu A^\mu +A_\mu J^\mu \right),
\end{equation}
where we have coupled the Proca field to a source $J^{\mu}$. While gauge invariance is broken by the explicit mass term, we can restore it through the introduction of a new field which is a gauge redundancy and thus does not change the physical theory. We generate an equivalent action for the Proca theory by the shift,
\begin{equation}
    A_\mu\rightarrow A_\mu+\partial_\mu\vartheta
\end{equation}
Note that this is not a gauge transformation {\it per se}, since gauge invariance has already been broken by the mass term. The action becomes, 
\begin{equation}\label{massivestuk}
S=\int d^4x \left( -\frac{1}{4} F_{\mu\nu}F^{\mu\nu}-\frac{1}{2} m^2 (A_\mu+\partial_\mu\vartheta)^2+A_\mu J^\mu-\vartheta\partial_\mu J^\mu\right).
\end{equation}
We expand the term $(A_\mu+\partial_\mu \vartheta)^2$ and rescale $\vartheta= \phi/m$ to normalize the $\vartheta$ kinetic term, leading to
\begin{equation}\label{massivephoton}
S=\int d^4x\left( -\frac{1}{4} F_{\mu\nu}F^{\mu\nu}-\frac{1}{2}m^2 A_\mu A^\mu -m A_\mu\partial^\mu\phi -\frac{1}{2}\partial_\mu\phi\partial^\mu\phi+A_\mu J^\mu-\frac{1}{m}\phi \partial_\mu J^\mu \right),
\end{equation}
This action has a gauge symmetry given by,
\begin{equation}
\delta A_\mu=\partial_\mu \Lambda,\ \ \ \delta\phi=-m\Lambda.
\end{equation}
We can now understand the massless limit: Provided that the current $J^\mu$ is conserved, $\partial_\mu J^\mu=0$, the Lagrangian becomes
\begin{equation}{
\lim_{m\rightarrow 0}\mathcal{L}= -\frac{1}{4} F_{\mu\nu}F^{\mu\nu} -\frac{1}{2}\partial_\mu\phi\partial^\mu\phi+A_\mu J^\mu,}
\end{equation}
and the gauge symmetry is 
\be 
\delta A_\mu=\partial_\mu \Lambda,\ \ \ \delta\phi=0.
\ee
Indicating that the massless limit of the Proca theory is standard U(1) gauge theory along with a scalar field $\phi$ which is completely decoupled.

\subsubsection{Massless Limit of Massive Kalb-Ramond}

Now we consider the massless limit of the massive KR field. Applying the same Stueckelberg trick as above, we take the action $S_{\kr}$, Eq.~\eqref{eq:SKR}, and shift the field by,
\begin{equation}
    B_{\mu \nu}\rightarrow B_{\mu \nu} + f_{\mu \nu}
\end{equation}
where $f_{\mu \nu}=\partial_\mu a_{\nu} - \partial_\nu a_{\mu}$. The action becomes,
\begin{equation}
      S = \frac{1}{12}\int d^4x \left[H_{\mu\nu\rho}H^{\mu\nu\rho} - 3m^2 (B_{\mu\nu} + f_{\mu \nu} ) ( B^{\mu\nu} + f^{\mu \nu} ) +  g ( B_{\mu\nu} + f_{\mu \nu} ) j^{\mu \nu}\right], 
      \label{eq:S_KR_F}
\end{equation}
where we now explicitly couple the KR field to a current $j^{\mu \nu}$ that we take to be conserved in the massless limit, i.e. $\lim _{m\rightarrow 0}\partial_\mu j^{\mu \nu}=0$. Expanding and canonically normalizing $m a_{\nu}=A_\nu$, we arrive at
\begin{align}
S =  \int  d^4 x & \left( \dfrac{1}{12}H_{\mu\nu\rho}H^{\mu\nu\rho}-\dfrac{m^2}{4}B_{\mu\nu}B^{\mu\nu} - \dfrac{m}{2}B_{\mu\nu}F^{\mu\nu} - \dfrac{1}{4}F_{\mu\nu}F^{\mu\nu} + g B_{\mu\nu}j^{\mu\nu} \right. \nonumber \\
  & + \left. \dfrac{2g}{m}A_\mu \partial_\nu j^{\mu\nu} \right)
\end{align}
where $F_{\mu \nu}=\partial_\mu A_{\nu} - \partial_\nu A_{\mu}$, and we have integrated by parts the coupling of $F_{\mu \nu}$ to the current $j^{\mu \nu}$. This theory has two distinct gauge invariances: the first given by the combined transformation,
\begin{equation}
    \delta B_{\mu \nu} = \partial_{\mu} \varepsilon_\nu - \partial_{\nu} \varepsilon_\mu \,\, , \,\, \delta A_{\mu} = - m \varepsilon_{\mu} ,
\end{equation}
which can most directly seen from Eq.~\eqref{eq:S_KR_F}, and a second separate gauge invariance involving only $A_\mu$
\begin{equation}
    \delta A_{\mu} = \partial_{\mu} \Lambda .
\end{equation}
In the massless limit, the theory becomes,
\begin{equation}
    \lim_{m\rightarrow0} S = \int d^4x \left( \frac{1}{12} H_{\mu\nu\rho}H^{\mu\nu\rho} -\frac{1}{4}F_{\mu \nu} F^{\mu \nu} + g B_{\mu \nu}j^{\mu \nu}\right)
\end{equation}
showing that the Stueckelberg vector field $A_{\mu}$ decouples in the massless limit, analogous to the scalar  Stueckelberg  of the Proca theory. This is qualitatively different from the massless limit of a vector field, wherein a Stuckelberg  scalar decouples in the massless limit. Recall that as in the Proca theory, if the current $j^{\mu\nu}$ is not conserved, then one cannot take the massless limit in a well-defined way, as $A_\mu$ becomes strongly coupled to the source \cite{Hinterbichler:2011tt}.

If we turn off the interaction, $g\rightarrow 0$, we can take this one step further and dualize the massless non-interacting KR field to an axion, leading to
\begin{equation}
    \lim _{\substack{g\rightarrow 0 \\ m\rightarrow0 }}S = \int d^4x \left( \frac{1}{2} (\partial \theta)^2 - \frac{1}{4}F_{\mu \nu} F^{\mu \nu} \right).
\end{equation}
where $\theta$ is a pseudo-scalar. This differs from the the massless non-interacting limit of Proca theory, wherein the spin-0 degree of freedom is a scalar. Thus {\it even in the massless non-interacting limit, the Proca and KR theories are physically inequivalent}; being distinguished the parity of their scalar degree of freedom. Furthermore, as shown above, the theories differ as soon as interactions are turned on. While the vector $A_{\mu}$ decouples from sources in the massless limit of Kalb-Ramond $B_{\mu \nu}$, in the Proca case it is the scalar $\phi$ that decouples from sources.

\section{Interactions of KRLPs\label{sec:Interactions}}

We now consider the interacting massive KR theory. 
We primarily take an effective field theory approach, with the interactions dictated by the symmetry properties of the KR field, though we will also discuss the string theory origins of these couplings. Recall that we delineate the  four-dimensional effective field theory of a rank-2 tensor from the string theoretic Kalb-Ramond field by referring to the former as KRLP. In this section we remain agnostic as to the UV completion of the massive KRLP, and in the next section we will discuss the emergence of KRLPs in string theory.\footnote{We note that for the interactions discussed in this section, there is no known field-theoretic UV completion in four dimensions. The UV complete theory would be string theory, which in the conventional case of critical string theory, requires extra dimensions.}

Outside of string theory, KRLPs can arise in a variety of effective field theory contexts. For example, in the Standard Model of particle physics, antisymmetric two-form fields appear as a dual description of pseudovector mesons \cite{Terschlusen:2013iqa}. KRLPs can also appear in the context of extra dimensions, including string theory, from Kaluza-Klein reduction of 2-form or higher-rank forms, e.g., a three form $C_{\mu \nu \rho}$, dimensionally reduced on a circle \cite{Copeland:1984qk}. Additionally, in condensed matter physics, KRLPs describe the interactions of vortices \cite{Beekman:2010zx}. This is analogous to the situation with ALPs, which naturally arise in string theory but are not resticted to that context;  an axiverse can instead emerge in dark QCD from the spectrum of dark pions \cite{Alexander:2023wgk,Maleknejad:2022gyf}.

To generically enumerate the interactions of KRLPs, we consider its following symmetry properties:
\begin{itemize}
    \item Antisymmetry: The KRLP $B_{\mu \nu}$ is an antisymmetric matrix.
    \item Parity: Unlike the electromagnetic potential $A_{\mu}$, which, as a vector, is parity-odd, the KRLP, $B_{\mu \nu}$, is parity-even, making it a pseudo-vector. This makes the field strength $H_{\mu \nu \sigma}$ parity-odd, and the dual field strength $(\star H)_\mu$ parity-even, again a pseudo-vector. This is consistent with the identification of the KR axion via $\partial_\mu \phi \equiv (\star H)_\mu $ in the massless limit, and is required for the KR gauge transormation $B_{\mu \nu}\rightarrow B_{\mu \nu} + F_{\mu \nu}$ to have definite parity.
\end{itemize}
These considerations suggest the following portals to the standard model:
\begin{itemize}
    \item {\bf Dark photon-like portal}: Analogous to the coupling of a gauge field $A_\mu$ to a fermion vector current  $j^\mu$, one can consider direct coupling of the potential $B_{\mu \nu}$ to a fermion current, $\bar{\psi}[...]\psi$ with an appropriate insertion of $\gamma$ matrices $[..]$. The antisymmetry and parity properties of $B_{\mu \nu}$ selects out $j^{\mu \nu}\equiv -i\bar{\psi} \sigma^{\mu\nu} \psi$, with $\sigma^{\mu \nu}\equiv \frac{i}{2}[\gamma^\mu,\gamma^\nu]$ as the correct fermion current. Hence, the natural coupling to fermions is given by the dimension-4 operator
    \begin{equation}
    \label{eq:3.1}
        {\cal L} = -i g B_{\mu \nu} \bar{\psi} \sigma^{\mu \nu}\psi
    \end{equation}
     where $g$ is a dimensionless coupling parameter and $\psi$ is a SM fermion field. This coupling is additionally motivated by the KRLP gauge transformation which mixes the KRLP with the photon field strength, $B_{\mu \nu} \rightarrow B_{\mu \nu} + F_{\mu \nu}$. The latter can also couple to fermions via the magnetic dipole coupling to yield the dimension-5 operator ${\cal L}\sim F_{\mu \nu}\bar{\psi} \sigma^{\mu \nu}\psi$, which is well studied in the context of dark matter \cite{Krnjaic:2022wor}. 
     EFT arguments aside, a possible string theory origin of this coupling was given in \cite{Dick:2020eop}, and aspects of the resulting phenomology were studied in \cite{Dashko:2018dsw}.

    It is instructive to compare the Kalb-Ramond dark photon-like portal with the usual dark photon scenario. While both feature a direct coupling of the gauge potential to a fermion current, the dark photon coupling necessarily emerges from a kinetic mixing with the SM photon, ${\cal L} = \varepsilon (F_{\mu \nu})^{\rm (DM)} (F^{\mu \nu})^{\rm (SM)}$, which, after diagonalization of the kinetic terms, generates the `millicharge' couplings $\varepsilon(A_\mu)^{\rm (DM)} (j^{\mu})^{\rm (SM)}$ as well as a mass mixing $\varepsilon (A_{\mu})^{\rm (DM)} (A^{\mu}) ^{\rm (SM)}$. The consequent mass mixing leads to the characteristic phenomenon of photon-dark photon oscillations, a feature which is absent for the KRLP. This is due to the fact that the analogous KRLP-photon mixing term, $B_{\mu\nu}F^{\mu\nu}$, is non-diagonalizable. Thus, for the KRLP, the coupling, $g$, is a free parameter arising directly from the coupling of KRLPs to fermions, rather than from kinetic mixing with the SM photon.

    An additional distinction with regards to the dark photon lies in the temperature dependence of the coupling to the fermion current: whereas the coupling of Kalb-Ramond $g$ is independent of the temperature of the SM thermal bath, the dark photon coupling $\chi$ is extremely sensitive to temperature. This is due to photon-dark photon conversion generated by the mixing term, which leads to a resonance as the temperature falls below the mass of the dark photon. This resonance is a primary physical effect in freeze-out of dark photons and dark photon mediated dark matter, as discussed in e.g. \cite{Feng:2017drg}, and is absent for the KRLP field.
     
    \item {\bf Axion-like portal}: In the massless limit, the Kalb-Ramond field is dual to an axion via the identification $(\star H)_\mu = \partial_\mu \theta$.\footnote{As discussed above, $(\star H)_\mu$ is a pseudovector, and is therefore parity-even. This ensures that $\phi$ is a pseudoscalar, which is necessary for its identification as an axion-like particle.} A natural expectation for the {\it massive} KRLP is that the interactions conventionally associated with $\partial_\mu \theta$, such as the coupling to the axial vector current $j^{\mu5}$, are lifted to couplings of $(\star H)_\mu$ which correctly reproduce the couplings of $\phi$ in the massless limit. To this end, we focus our attention on the interaction:
    \begin{equation}
    \label{eq:4.2}
    {\cal L} = \tilde{g} \tilde{H}_\mu\bar{\psi} \gamma^\mu\gamma^5 \psi,    
    \end{equation}
    where $\tilde H\equiv \star H$, and $\tilde{g}$ is a coupling constant with mass dimension $-1$, similar to the axion-electron coupling $g_{aee}$ or the axion-photon couping $g_{a\gamma\gamma}$.

    This coupling emerges from a simple shift in the KR field strength, in a manner consistent with the KR symmetries, by altering the kinetic term to
    \begin{equation}
    \label{eq:4.3}
        S = \int {\rm d}^4 x \sqrt{-g}\,(H_{\mu \nu \sigma} + \tilde{g}\epsilon_{\mu \nu \sigma \rho}j^{\rho5})(H^{\mu \nu \sigma} + \tilde{g}\epsilon^{\mu \nu \sigma \rho}j_{\rho5}),
    \end{equation}
    where $j^{\mu5}$ is the axial vector current. The cross term $\epsilon_{\mu \nu \sigma \rho} j^{\rho 5}H^{\mu \nu \sigma}$ is precisely Eq.~\eqref{eq:4.2} given above.

    In the context of string theory, axion-like couplings of the KR field arise in the context of anomaly cancellation (see \cite{Polchinski:1998rr} for a textbook review), wherein the KR field strength appears in the combination,
    \begin{equation}
        S \sim \int (H_3 + \alpha'\Omega_3)\wedge \star (H_3 + \alpha'\Omega_3)
    \end{equation}
    where $\Omega_3$ is a Chern-Simons three-form. The cross term $\Omega_3 \wedge \star H $ generates axion-like couplings. For example, in the massless limit, $\star H = d\theta$, and in the case of an Abelian gauge field $\Omega_3 =A \wedge dA$, one finds $ \int  \Omega_{A} \wedge d\theta \sim - \int d^4x \theta F_{\mu \nu} (\star F)^{\mu \nu}$; the conventional axion-photon coupling.  To realize the coupling to fermions, one can replace $\Omega_3$ in the above with $\star j^{5}$ with $j^5 \equiv j_{\mu 5}{\rm dx}^\mu$ with $j^{\mu 5}$ the axial vector current; in this case one finds $S \sim \int \star H \wedge \star j^5 \sim \int d^{4}x \tilde{H}_\mu j^{\mu 5}$, reproducing Eq.~\eqref{eq:4.2}
\end{itemize}

The interactions Eqs.~\eqref{eq:3.1} and \eqref{eq:4.2} will be the focus of our cosmological analyses. For completeness, and to bolster the motivation for this setup, we first take a detour through the string theory origins of antisymmetric tensor fields such as the Kalb-Ramond field, including the mass and interactions of the 4d theory.

\section{String Theory Origins } 
\label{Sec:Strings}
To inform our analysis of the interacting massive Kalb-Ramond theory, we return to the historical origin of the KR field, which lies in string theory \cite{Kalb:1974yc}. Indeed, the KR field describes the generalization of electrodynamics from a theory point particles to that of strings; while the path through spacetime of a point particle is a line (a {\it worldline}) and the interaction with electromagnetism is the integral over the worldline of a 1-form, namely the electromagnetic potential $A$, a string instead traces out a surface in spacetime, namely a {\it worldsheet}, and the corresponding electromagnetic-like interaction is an integral over the worldsheet of a two form $B_2$, namely the KR field.

In this section we provide a brief discussion of the string theory description of the KR field and the resulting four-dimensional physics. For this discussion it will be convenient to work with differential form notation, and to this end express the KR field as a two form,
\begin{equation}
B _2= B_{\mu \nu} {\rm d}x^\mu \wedge {\rm d}x^\nu.
\end{equation}
The component field $B_{\mu \nu}$ is antisymmetric in $\mu$ and $\nu$, by construction. The interaction with a string is simply given by,
\begin{equation}
\label{eq:string}
S_{\rm string}\sim
\int _{\Sigma^2} B_2
\end{equation}
where $\Sigma^2$ is the worldsheet. In component form this reads,
\begin{equation}
S_{\rm string} \sim 
\int d^2 \sigma \epsilon^{a b} \, \partial_a X^\mu \partial_b X^{\nu} B_{\mu \nu}
\end{equation}
$X^\mu$ are a set of scalar fields that give the position of the worldsheet as it propagates through spacetime.

Quantization of the string and cancellation of all anomalies leads to five consistent string theories \cite{Polchinski:1998rr}. At leading order in the string coupling, these can each be described by supergravity in 10 dimensions \footnote{Intrinsically four-dimensional string theories, in contrast to the 10 dimensions of (critical) Type II, Type I, and Heterotic, have been proposed in \cite{Gates:1987sy,DEPIREUX1989364,Gates:1989yw,DEPIREUX1990411,BELLUCCI198967}}.  Here we focus on Type IIB string theory, since this is the context for most modern string phenomenology; see, e.g., \cite{Brandenberger:2023ver,Cicoli:2023opf} for recent reviews of string cosmology.  The action of the bosonic sector of type IIB supergravity is given by \cite{Polchinski:1998rr},
\begin{equation}
S_{\rm IIB, bosonic} = S_{\rm NS} + S_{\rm R} + S_{\rm CS}.
\end{equation}
The Neveu-Schwartz sector (``NS'') is given by 
\begin{equation}
S_{\rm NS} = \frac{1}{2\kappa_{10}^2}\int {\rm d}^{10}x (- G)^{1/2} e^{- 2 \Phi} \left( R + 4 \partial_\mu \Phi \partial^\mu \Phi - \frac{1}{2}|H_3|^2\right),
\end{equation}
where $|H_3|^2 \equiv H_{\mu \nu \sigma}H^{\mu \nu \sigma}$ where $H$ is the Kalb-Ramond field strength, and $\Phi$ is the dilaton field.  The Ramond-Ramond (R) sector contains three copies of electrodynamics, one of a scalar potential $C_0$,  one with a two-form potential $C_2$, and one copy with a four-form potential $C_4$, with corresponding field strengths $F_1$, $F_3$, and $F_5$. The Ramond sector action is given by
\begin{equation}
\label{eq:SR}
S_{\rm R}=\frac{1}{2\kappa_{10}^2}\int {\rm d}^{10}x (- G)^{1/2}  \left( |F_1|^2 + |\hat{F}_3|^2 + \frac{1}{2}|\hat{F}_5|^2 \right),
\end{equation}
where $\hat{F}_3$ and $\hat{F}_5$ are the {\it corrected } field strengths, which include a $B_2$-dependent shift:
\begin{equation}
\hat{F}_3 = F_3 - C_0 \wedge H_3,
\end{equation}
\begin{equation}
\hat{F}_5 = F_5 - \frac{1}{2}C_2 \wedge H_3 + \frac{1}{2}B_2 \wedge F_3.
\label{eq:F5}
\end{equation}
Finally, the Chern-Simons (CS) action describes the coupling between NS and R sectors, given by
\begin{equation}
S_{\rm CS} = -\frac{1}{4 \kappa_{10}^2}\int C_4 \wedge H_3 \wedge F_3.
\end{equation}
From this one may appreciate that there is a rich set of interactions between the Kalb-Ramond field and the other $p$-form fields of string theory.

The Kalb-Ramond field also interacts with extended objects in string theory, e.g., D-branes of varying dimensions \cite{Johnson:2000ch}. This is described in part by the Dirac-Born-Infeld (DBI) action, which for a $p+1$-dimensional (``Dp'') brane is given by
\begin{equation}
S_{\rm Dp, DBI} = - T_p \int {\rm }d^{p+1} \sigma \, \sqrt{ -  {\rm det} (h_{\mu \nu} + \partial_\mu \Phi^a \partial_\nu \Phi_a  + F_{\mu\nu} + B_{\mu \nu }}) .
\end{equation}
Here  $T_p$ is the brane tension, $\sigma$ are coordinates on the brane, and $h_{\mu \nu}$ is the metric on the brane. The set of scalars $\Phi^a$  correspond to the position of the brane, and the quantized excitations of each scalar physically correspond to fluctuations of the brane along each coordinate direction. Similarly, the worldvolume gauge field, with strength $F_{\mu \nu}$, corresponds to vector-like fluctuations of the shape of the brane.

Expanding the DBI action in small field fluctuations generates an explicit mass term for $B_{\mu \nu}$ along with kinetic terms of the scalars and gauge field, and a wealth of interactions between the KR field $B_{\mu \nu}$ and the metric $h$, the scalars $\Phi$, and the gauge field $A_\mu$. These can be enumerated via an expansion of the DBI action, e.g. the KR mass term and kinetic terms for the scalar and gauge field each arise at quadratic order, as
\begin{equation}
    S_{\rm Dp, DBI} \supset - T_p \int {\rm }d^{p+1} \sigma \,\; \sqrt{-h}\left[ B_{\mu \nu}B^{\mu \nu}+ \frac{1}{2} |\partial \Phi \partial \Phi|^2 + \frac{1}{4}F_{\mu \nu} F^{\mu \nu} + .... \right].
\end{equation}
Extending the D-brane action to include thre worldvolume fermions \cite{Dasgupta:2016prs,Marolf:2003vf,Martucci:2005rb,Grana:2002tu,Tripathy:2005hv} leads to yet more interactions involving the Kalb-Ramond field.

Branes also interact with Kalb-Ramond field via the Chern-Simons action that describes the charge of the brane under various form fields. This is given by,
\begin{equation}
\label{eq:4.12}
S_{\rm Dp-CS} = \int \sum_{p=q+r} C_q \wedge (e^{B+F})_r
\end{equation}
where $C_q$, $q=0,2,4$, are the Ramond-Ramond potentials, and $(e^{\cal X})_r$ denotes the $r$-form term in the series expansion of $e^{\cal X}$. For example, in the case of a D3-brane, one has 
\begin{equation}
\label{eq:4.13}
S_{\rm D3-CS} =\int C_4 + \int C_2 \wedge (B + F )  + \int C_0 (B + F )\wedge (B+F).
\end{equation}

From this one can see that a D3 brane in vacuum is electrically charged under $C_4$ field, whereas a D3 brane in a background B-field also couples to $C_2$ and $C_0$. Conversely, the CS action encodes interactions of $B_2$ with the form fields $C_{0}$, $C_2$, and $C_4$, as well as the worldvolume gauge field strength $F$.

\subsection{From 10 dimensions to 4 dimensions}

The effective action describing the four-dimensional Kalb-Ramond field is derived from Kaluza-Klein reduction of the ten-dimensional theory. At lowest order, we can expand the KR field as \cite{Grimm:2004uq}:
\begin{equation}
\label{eq:B2dimred}
B _{\mu \nu}(x,y)  = \begin{cases}
    B_{\mu \nu} ^{(4{\rm d})} (x) & \text{for $\mu ,\nu= 0,1,2,3$ } \\
   \sum_I \phi^I (x) \omega ^I _{\mu \nu} (y)& \text{for $\mu,\nu=4...9 $} 
  \end{cases}
\end{equation}
where $x$ and $y$ are the coordinates on the 4d and 6d internal space respectively, $B_{\mu \nu} ^{(4{\rm d})} (x)$ is the 4d KR field (we henceforth drop the $(4{\rm d})$ superscript), and $\phi^I$ are a set of psuedoscalar (``moduli'') fields and $\omega ^I $  are a set of harmonic two-forms in the internal space. The fields $\phi^I$ are axion-like particles, which can be useful for model building, but the primary focus of the current work is instead the 4d KR tensor field $B_{\mu \nu}$.

The action for the 4d KR field follows from inserting the expansion above in the 10d action. For example, dimensional reduction of the kinetic term proceeds as
\begin{equation}
S_{\rm kinetic} = \int {\rm d}^{10}x \sqrt{-G} |H_3|^2  \rightarrow {\rm Vol}(X)\int {\rm d}^{4}x \sqrt{-g}  |H_3|^2,
\end{equation}
where ${\rm Vol}(X)$ is the volume of the internal space and $g$ is the 4d metric. Dimensional reduction of other terms in the action proceeds similarly, leading to a mass and interactions of the 4d Kalb-Ramond field.

\subsection{Mass Generation Mechanisms of the Kalb-Ramond Field}

Since its inception in the 1970s, the Kalb-Ramond field has often been thought of as massless, leading to association with the `Kalb-Ramond Axion.' However, in the modern era of warped flux compactifications, the 4d Kalb-Ramond field is in general {\it massive}. This fact is widely utilized in cosmological models that utilize the $B_2$ and $C_2$ axions; the breaking of the KR gauge symmetry generates a mass for the 4d moduli fields (Kalb-Ramond `axions'), which is encoded in the 4d supergravity theory via the Kahler potential for the 4d scalar fields \cite{Cicoli:2023opf,Cicoli:2023qri,McDonough:2022pku}
\begin{equation}
K = -3 \ln \left({\rm Re}(T) - \frac{2\gamma}{g_s^2}\,b^2\right).
\label{NewK}
\end{equation}
where $T$ is the volume modulus and $b$ is a Kalb-Ramond axion arising from dimensional reduction. This generates a mass for $b$ via the scalar potential, $V = e^K \left( |DW|^2 - 3 |W|^2\right)$ where $W$ is the superpotential.

Based on this, one might expect the 4d KR tensor field $B_{\mu \nu}$ to also be massive. Indeed, compactification on an O3/O7 Calabi-Yau orientifold, useful for breaking ${\cal N}=2 \rightarrow 1$ supersymmetry, generates a high-scale mass for both $B_2$ and $C_2$, while compactification on an O5/O9 generates a mass for $B_2$ while leaving the $C_2$ massless in four dimensions (c.f. Table 5.6 in \cite{Grana:2005jc}). Here we focus on features of the ${\cal N}=2$ theory in 10 dimensions, without specific reference to the compactification (e.g., on an orientifold as in KKLT). As we will show, already at this level the KR mass receives contributions from multiple sources, both perturbative and non-perturbative, and arising from both the bulk supergravity action and the D-brane action.

 As a simple example, one may appreciate from Eq.~\eqref{eq:SR} that a mass for the four-dimensional $B_{\mu \nu}$ will be generated upon dimensional reduction whenever there is a background $F_3$ flux in the internal space, due to the $B_2$ implicitly appearing in $\hat{F}_5$. More concretely, a mass in four dimensions may be generated from ten dimensions via the term
\begin{equation}
    |\hat{F}_5|^2  \supset B_{\mu \nu} B^{\mu \nu} (F_3 )_{ mnp} (F_3) ^{mnp}.
\end{equation}
where $\mu,\nu$ and $m,n,p$ denote 4d $(0,1,2,3)$ and internal space $(4...9)$ indices respectively.

For example, if one considers a background $F_3$ flux given by $f_3 \equiv \int_{\Sigma_3} F_3$, where $\Sigma_3$ is a three-cycle in the internal space, then dimensional reduction leads to a mass for the 4d KR field, $m _{\rm KR} \propto f_3 $. Such fluxes are a generic feature of the standard frameworks for moduli stabilization in string theory, such as KKLT \cite{Kachru:2003aw} and the Large Volume Scenario (LVS) \cite{Balasubramanian:2005zx, Cicoli:2008va}, which both rely on the flux-induced superpotential to stabilize the complex structure moduli.

A mass for the KR field is also generated by interactions with D-branes. For example, consider the quadratic term in the expansion of the DBI action,
\begin{equation}
    {S}_{\rm DBI} \supset - T_p \int {\rm }d^{p+1} \sigma \, \sqrt{-h} \;  B_{\mu \nu}B^{\mu \nu}.
\end{equation}
Reducing to four dimensions, one finds
\begin{equation}
    {S}_{\rm DBI} ^{\rm 4d} \supset - \int {\rm d^4}x \sqrt{-g} \, \left[ T_p {\rm Vol}(\Sigma) \right] \, B_{\mu \nu} B^{\mu \nu},
\end{equation}
where $\Sigma$ is the cycle in internal space that is wrapped by the brane. The non-vanishing size of the cycle wrapped by the brane, itself due to the outward pressure exerted by fluxes, leads to additional contribution to the KR mass, as $m_{KR}^2 \propto T_p {\rm Vol}(\Sigma)$ .

D-branes can also generate a mass via the presence of fluxes on brane worldvolume, namely when the worldvolume gauge field strength is non-vanishing, $f_2 = \int_{\Sigma^2} F_2 \neq 0 $ , where $\Sigma^2$ is a two-cycle wrapped by the brane. 
This generates a mass for the 4d Kalb-Ramond field from dimensional reduction of the quartic interaction,
\begin{equation}
S_{\rm DBI} \supset \int d^{p+1}x\, \sqrt{-h} F_{mn}F^{mn} B_{\mu\nu}B^{\mu \nu} \rightarrow \int {\rm d^4}x \sqrt{-g} f_2 ^2 B_{\mu \nu}B^{\mu \nu}
\end{equation}
corresponding a mass for the 4d Kalb-Ramond field
proportional to the total amount of worldvolume fluxes.

Finally, the Kalb-Ramond mass also receives non-perturbative contributions. A prototypical example is that generated by gaugino condensation on a stack of D7 branes, the latter a crucial component in both the KKLT and LVS scenarios for stabilizing the volume moduli. Recall that the brane itself is host to not just a gauge field $A_\mu$, but to a supersymmetric theory, which includes superpartner(s) to the gauge field, namely the gaugino(s). In the presence of $N$ multiple coincident branes, the theory is an $SU(N)$ super-Yang Mills theory \cite{Polchinski:1998rr},  the vacuum of which is characterized by a non-vanishing fermion bilinear, $\langle \lambda \lambda \rangle \neq 0$ \cite{Affleck:1983mk}, referred to as a gaugino condensate. The study of this in a KKLT-type background has been a subject of intense study \cite{Kallosh:2019axr,Kallosh:2019oxv,Carta:2019rhx,Kachru:2019dvo,Gautason:2019jwq,Hamada:2019ack,Hamada:2018qef}. If we instead consider a background with a KR field in the 4d spectrum, the expansion of the DBI action on the D7-brane stack would lead to a Kalb-Ramond mass $m_{\rm KR}$ proportional to the condensate VEV $\langle \lambda \lambda \rangle$.

\subsection{Kalb-Ramond-Like Particles (KRLPs) from String Theory}

This work does not focus on the Kalb-Ramond field of string theory, but instead the more general possibility of a Kalb-Ramond-Like-Particle (KRLP). Here we brielfy examine the possibility of KRLPs in string theory.

As a simple warm-up, one may consider a toy model in five dimensions, where the fifth dimension is a circle of radius $R$. Proposed in \cite{Arkani-Hamed:2003xts}, this is a simple playground for understanding ALPs: a 5d gauge field $A_{\mu}$ leads to a 4d ALP $\phi (x) \equiv A_5$ upon dimensional reduction (where 5 denotes the compact direction). An equally simple construction is a rank-3 antisymmetric tensor field $C_{\mu \nu \rho}$ in five dimensions, such as arises in Type IIA string theory. In this case dimensional reduction leads to a 4d KRLP, $B_{\mu \nu}\equiv C_{\mu \nu 5}$. This easily generalizes to more compact directions, e.g., $C_{\mu \nu \rho}$ on an N-torus $T^N$ would produce $N$ number of KRLPs. 

From this simple example, one may appreciate that Type IIA string theory compactified on $T^6$ is a direct path to 6 KRLPs. We leave a more comprehensive analysis of KRLPs in Type IIA to future work.

In the context of a Calabi-Yau compactification of Type IIB string theory, a KRLP can arise from dimensional reduction of the $C_2$ Ramond-Ramond field.  Dimensional reduction of $C_2$ follows the same procedure as that for $B_2$,   Eq.~\eqref{eq:B2dimred}, and can give rise to a single KRLP in four dimensions. Unlike the fundamental Kalb-Ramond $B_2$, perturbative corrections to IIB supergravity do not generate a mass for $C_2$. This is similar to $C_2$ axions (i.e., the scalar moduli that arise from dimensional reduction), which makes them an excellent candidate for axion dark  matter or dark energy \cite{Cicoli:2023qri}. The mass of the the $C_2$ KRLP $C_{\mu \nu}$ comes instead purely from non-perturbative effects, such as the DBI-like action on the worldvolume of NS5 branes \cite{Barnaby:2011qe}, or else from an O3/O7 orientifolding. 

We leave a detailed study of the string theory KRLPs and their phenomenology, in particular in compactifications other than Type IIB Calabi-Yau orientifolds, for future work. We emphasize here that, analogous to the relation between the axion and ALPs, antisymmetric rank-2 tensor fields are not limited to the field originally proposed by Kalb and Ramond in \cite{Kalb:1974yc}. Absent an explicit string theory construction, and in the spirit of EFT, in what follows we remain agnostic as to the range of mass and couplings of KLRPs, and to their UV completion, string theory or otherwise.

\section{Freeze-In Production of KRLPs}

\label{sec:freezein}

To initiate the study of Kalb-Ramond cosmology, we study the freeze-in production of KRLPs generated by the interactions discussed in Sec.~\ref{sec:Interactions}. We begin with the a review of the freeze-in mechanism \cite{Hall:2009bx}, and then specialize to the KRLP case.

\subsection{Dark Matter Freeze-In }

Cold dark matter can be produced non-thermally in the early universe via a process known as `freeze-in' \cite{Hall:2009bx}. In contrast to freeze-out, in which the dark matter begins in thermal equilibrium with the standard model plasma, in freeze-in production the dark matter was never in equilibrium with the thermal bath. Rather, it is produced via extremely weak interactions with SM particles. We will consider three different tree-level processes arising from the interactions discussed in Section~\ref{sec:Interactions}, shown in Table~\ref{tab:feynman-diagrams}. They are inverse decay or `coalescence,' fermion-antifermion annihilation, and Compton scattering.   
\begin{table}[ht!]
    \centering
    \begin{tabular}{|c|c|}
        \hline
        Process & Feynman Diagrams \\
        \hline
        Inverse Decay: $f\bar{f}\rightarrow B$ &\includegraphics[width=0.2\linewidth]{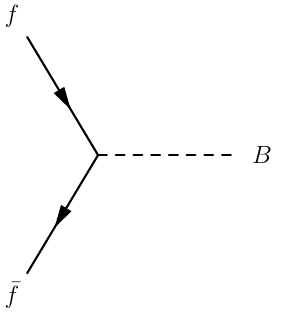} \\ \hline
         Annihilation: $f\bar{f}\rightarrow \gamma B$ &\includegraphics[width=0.45\linewidth]{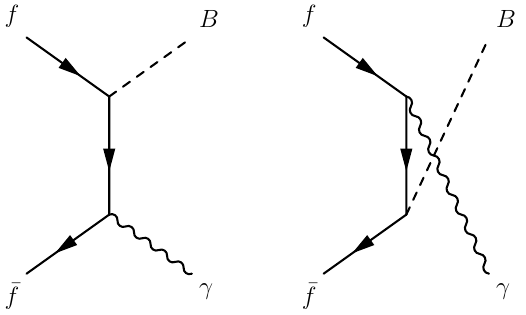} \\
         \hline
         Compton: $f\gamma\rightarrow f B$ & \includegraphics[width=0.45\linewidth]{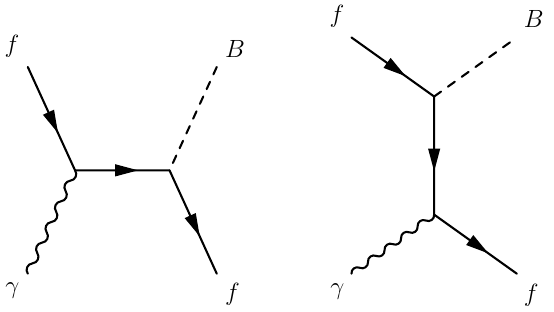} \\
         \hline 
    \end{tabular}
    \caption{Tree level diagrams contributing to the freeze-in of a KRLP.}
    \label{tab:feynman-diagrams}
\end{table}

We will find that the coalescence channel leads to production of particles which are unstable and thus will decay too quickly to account for the DM relic density. Thus, we will focus our attention on the annihilation and Compton channels.
The $2\to2$ freeze-in processes are $e^+e^-\to\gamma B$ (annihilation) and $e^\pm\gamma\to e^\pm B$ (Compton).  The evolution of the $B$ number density due to the process $12\to3B$ evolves in time as
\begin{align}
    \dot{n}_B +3Hn_B = \overline{n}_1 \overline{n}_2 \, \langle \sigma v \rangle_{12\to3B}
\end{align}
where $\overline{n}_i$ is the equilibrium number density for species $i$: $\overline{n}_i = (g_i/2\pi^2)Tm_i^2K_2(m_i/T)$ where $K_j$ is the modified Bessel function of order $j$ (assuming Boltzmann statistics).  The thermal average $\langle \sigma v \rangle_{12\to3B}$ is
\begin{equation}
\langle \sigma v \rangle_{12\to3B}=   \frac{ \bigints \! ds\,
    \sigma_{12\to3B}(s)\,K_1(\sqrt{s}/T)\, \sqrt{s}\left[ s-2(m_1^2+m_2^2) + \dfrac{(m_1^2-m_2^2)^2}{s} \right] } { 8T\,m_1^2K_2(m_1/T)\,m_2^2K_2(m_2/T) } \,.
\end{equation}
The upper limit on the integral is infinity and the lower limits on the integral are $s_\mathrm{MIN}=\mathrm{MAX}(4m_f^2,m_B^2)$ for annihilation and $s_\mathrm{MIN}=(m_f+m_B)^2$ for Compton. 

It is convenient to work in the dimensionless variable $x\equiv m_B/T$ to track the evolution with decreasing temperature. Noting that $x\propto T^{-1} \propto a$, we can re-write time derivatives in the following way:
\begin{equation}
    \dot{a}\frac{dn_B}{dx}+3Hn_B = \langle \sigma v \rangle_{12\to3B} \,\overline{n}_1\overline{n}_2,
\end{equation}
and make use of $H=\dot{a}/a$. After integrating by parts, and making use of the fact that the entropy density scales as $s=s_0 a^{-3}$, one arrives at:
\begin{equation}
    \frac{d Y_B}{dx}=\frac{\langle \sigma v \rangle_{12\to3B} \,\overline{n}_1\overline{n}_2}{x s(x) H(x)},
\end{equation}
where we've defined the dimensionless abundance $Y_B \equiv n_B/s$. We have assumed the number of relativistic species is not changing during the temperature range of interest. In terms of $x$, we have that
\begin{equation}
    s(x)=g_{\star}\frac{2\pi^2}{45}T^3=c_1 T^3 = c_1\left(\frac{m_B}{x}\right)^3,
\end{equation}
and
\begin{equation}
H(x)=\sqrt{\frac{g_{\star}\pi^2}{90}}\frac{1}{M_{\pl}}T^2 = \frac{c_2}{M_{\pl}} T^2 = \frac{c_2}{M_{\pl}}\left(\frac{m_B}{x}\right)^2 , 
\end{equation}
where $M_\pl$ is the Planck mass. From there, the fraction of the dark matter that the species of interest comprises can be estimated as:
\begin{equation}
    \Omega_B\approx \frac{m_B s_0}{\rho_{C}}Y^{\infty}_B \,,
\end{equation}
where $s_0$ is the present entropy density and $\rho_C=3H_0^2/8\pi G$ is the critical density.

\subsection{Kalb-Ramond Freeze-In Processes}
We now turn to study the freeze-in of KRLPs via the processes discussed above. We consider two distinct classes of interaction scenarios to determine the freeze-in relic abundance of KRLP dark matter, denoted as the `dark photon-like' interaction and the `axion-like' interaction, as discussed in Section \ref{sec:Interactions}.

\subsubsection{`Dark Photon-Like' Portal}
We first consider the `dark photon-like' interaction, Eq.~\eqref{eq:3.1}, in which the KRLP couples to SM fermions via a magnetic dipole-like interaction. The analogous dark photon interaction is that of a dark photon coupled to the fermion current: $\mathcal{L}_{\inter} \supset A_\mu' J^\mu$, where $A_\mu'$ is the dark photon,  or the dark photon field strength magnetic dipole interaction with the SM, $\mathcal{L}_\inter \supset F_{\mu\nu}' J^{\mu\nu}$. What we now have is a generalization of this interaction for the KRLP, in which the field itself couples to the tensor current $J^{\mu\nu}$. 

There are several key differerences to note between the interactions in the dark photon case and the KRLPs. First, the vector current $J^\mu$ is conserved, while the tensor current $J^{\mu\nu}$ is not. Furthermore, in the coupling to $J^{\mu\nu}$, the KRLP field couples directly to the current versus the field strength for the analogous dark photon scenario. We also reiterate a notable distinction between the KRLP interaction and that of a dark photon: for the usual dark photon scenario, the kinetic mixing between an SM photon and a dark photon accounts for thermal effects and thus the effective mixing parameter is temperature dependent  \cite{Redondo:2008aa, Redondo_2009}. For the KRLP dark photon-like portal, the coupling parameter between the KRLP and the SM fermions does not encode any temperature dependence.

\begin{figure}[ht!]
    \centering
    \includegraphics[width=0.99\linewidth]{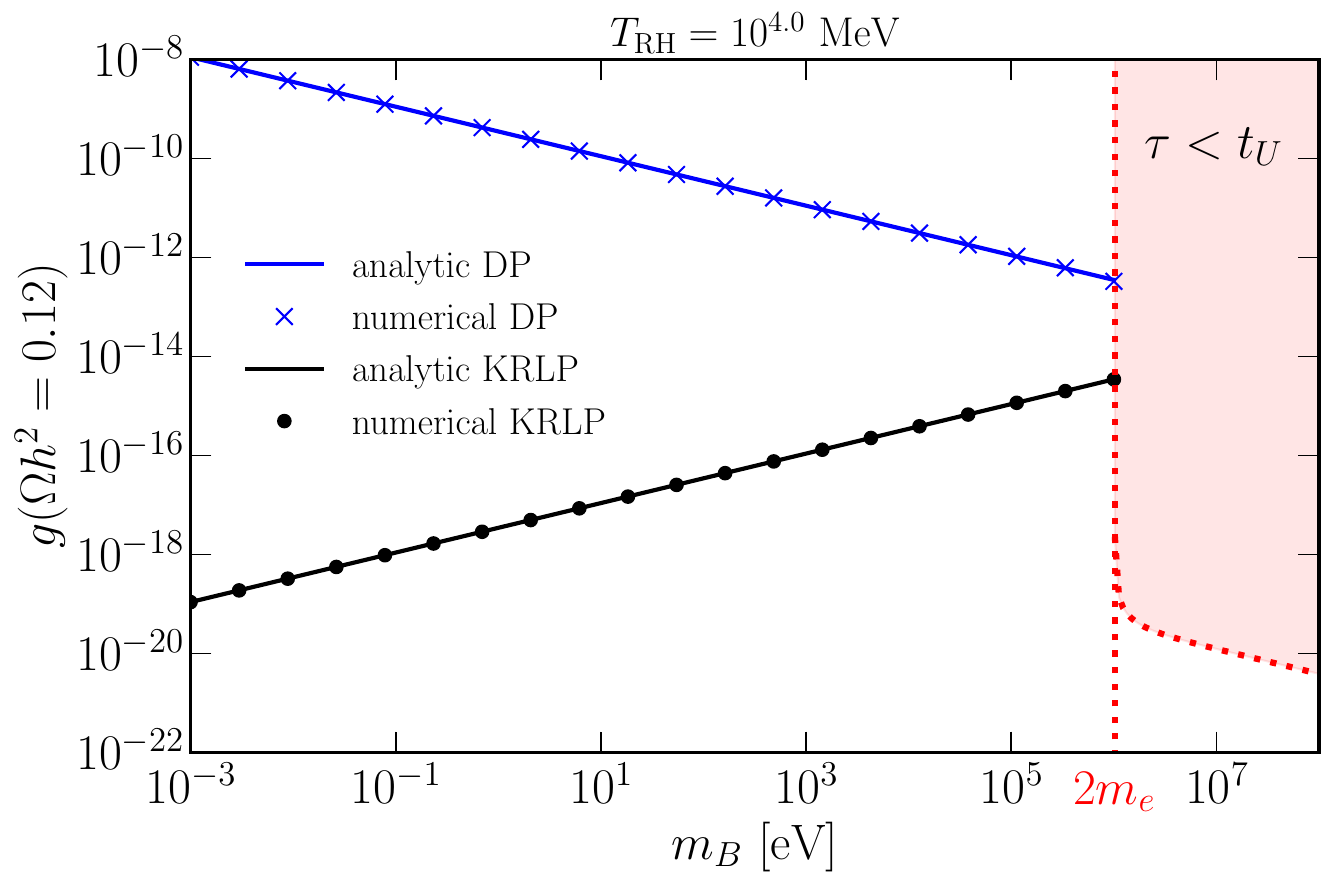}
   \label{fig:dipole-mass-scaling}
    \caption{Freeze-in via the Dark Photon-like Portal assuming electrons as the fermion in the processes. We show critical values of $g$ to result in $\Omega h^2=0.12$ for KRLPs (black) for $T_\mathrm{RH}=10^4 \mathrm{MeV}$, and for dark photons (blue). The vertical dashed line corresponds to the threshold mass $m=2m_e$, while the shaded red region denotes unstable relics with lifetimes $\tau < H_0^{-1}$. For KRLPs, the critical value of $g$ scales as $g=3.5\times10^{-16}(m_B/\mathrm{eV})^{1/2}(\mathrm{MeV}/T_\mathrm{RH})^{1/2}$ shown by the solid black  line; it is independent of $m_f$. For dark photons, the critical value of $g$ scales as $g = 3.5\times10^{-10}(\mathrm{eV}/m_B)^{1/2}(m_f/m_e)^{1/2}$ shown by the solid blue line; it is independent of $T_\mathrm{RH}$.   Values of $g(m_B)$ above the lines and below the values where freeze-in is operative lead to $\Omega h^2>0.12$ and hence are disallowed.   The values of $g$ in the figure are well within the region where freeze-in is operative. 
    }
    \label{fig:dipole-parameters}
\end{figure}

As stated above, from the interaction Eq.~\eqref{eq:3.1} we can obtain freeze-in production of KRLPs via three tree-level channels. The coalescence channel contributes to the relic density for masses $m_B > 2 m_f$, where $m_f$ is the fermion mass (this is in the instability range), whereas fermion-antifermion annihilation and Compton scattering contribute to the relic density for a larger mass range. We show the explicit cross section computations in Appendix~\ref{App:CrossSections}, and here just show the relic density results. Below, in Fig.~\ref{fig:dipole-parameters}, we show the parameter space for the mass and coupling for freeze-in via $2\rightarrow 2$ processes for the dark photon-like interaction for KRLPs in comparison to dark photons with the coupling $A_\mu' J^\mu$ to the SM. We show the coupling-mass parameter space for KRLP and DP dark matter where the SM fermion is an electron. The black curves denote the KRLP, while the dark photon results are shown in blue. We have assumed freeze-in, which requires $Y_B(\mathrm{freeze\ in})\ll \overline{Y}_B(\mathrm{freeze\ in})$, where the overbar denotes the equilibrium value (which is approximately $1$ when $x<1$). This places an upper limit on $g$. Note that formally, the dark photon coupling $g$ could be a function of $T$. However, this $T$-dependence does not change the overall scaling behavior for the dark photon, so for our illustrative purposes we simply take it to be a constant.

Results are shown in Fig.~\ref{fig:dipole-parameters}. We find that KRLP production via the dark photon-like coupling is extremely efficient at low masses,  ruling out a large window of parameter space $g\gtrsim 10^{-15}$ where KRLPs are overproduced. This effect is due to factors of $1/m_B^2$ which appear in the cross sections for the annihilation and scattering process. This is in contrast with the  the analogous dark photon processes, for which this mass dependence of the cross section vanishes, leading to the characteristic upward slope of seen in the purple curves in Fig.~\ref{fig:dipole-parameters}. The origin of this difference is that the vector current, $j^\mu$ is a conserved quantity, whereas $j^{\mu \nu}$ to which the Kalb-Ramond couples, is not. It is a straightforward exercise to demonstrate that the tensor current satisfies 
\begin{equation}
    \partial_\mu j^{\mu \nu} = - 4 m j^{\nu} - 2 \partial^\nu (\bar{\psi}\psi), 
\end{equation}
and thus is not conserved. This suggests that the momentum-dependent terms in the polarization sum will contribute to the cross section in contrast to the dark photon cross section. One may also see explicitly that the $1/m_B^2$ term in the polarization sum is not annihilated. We show this in detail for the case of Compton scattering in Appendix \ref{App:CrossSections}. As a result, we find that the scaling of the coupling, $g$, to obtain the critical value of $\Omega$ differs between the two cases. We see that for the KRLP,
\be 
g_\kr =3.5\times10^{-16}(m_B/\mathrm{eV})^{1/2}(\mathrm{MeV}/T_\mathrm{RH})^{1/2},
\label{eq:gkr}
\ee 
whereas for the dark photon, 
\be
g_\dph = 3.5\times10^{-10}(\mathrm{eV}/m_B)^{1/2}(m_f/m_e)^{1/2}. 
\label{eq:gdp}
\ee 
We show an explicit computation of the scaling in Appendix~\ref{App:scaling}.

From the scaling of $g$ in each of the two scenarios, one can appreciate further qualitative differences between the two. In addition to the couplings to electrons, which we have explicitly shown in Figure~\ref{fig:dipole-parameters}, one can also consider freeze-in via couplings to the muon or tau. In contrast to the dark photon case, in which increasing the fermion mass then in turn requires larger couplings to recover the same abundance, the KRLP abundance is \textit{independent} of the fermion mass, which can be seen explicitly from Eqs~\eqref{eq:gkr} and~\eqref{eq:gdp}. On the other hand, KRLP freeze-in is sensitive to the reheat temperature, $T_\rh$, while the dark photon is not. In Figure~\ref{fig:dipole-parameters}, we take a fiducial value for the reheat tempearture, $T_{\rh} = 10^{4}$ MeV. Changing $T_\rh$ will alter the KRLP curve by a factor of $T_{\rh}^{1/2}$, and leave the dark photon curve unaltered, per Eqs.~\ref{eq:gkr} and ~\ref{eq:gdp}.

On the higher end of the KRLP mass range, similarly to the dark photon, it is difficult to get stable relics with masses above threshold $m_B>2m_f$. The lifetime of a KRLP in this regime is:
\begin{equation}
    \tau^{-1}_{B\rightarrow \bar{f}f} = \frac{1}{16 \pi }\frac{|\mathcal{M}|^2}{2m_B}\sqrt{1-\frac{4m^2_f}{m_B^2}}= \frac{g^2 m_B}{8\pi}\left(1-\frac{4m_f^2}{m_B^2}\right)^{3/2}.
\end{equation}
For particle lifetimes shorter than the age of the universe $\tau \lesssim 1/H_0 $, not only is it difficult to get the right relic abundance, but the injection of energy due to decay can radically alter the thermal history. This region is shaded red in Fig. \ref{fig:dipole-parameters}.

On the opposite end of the mass range, we also see that for masses  $m < {\rm MeV}$ the KRLPs produced via freeze-in are effectively a `warm' dark matter component. They are produced with momentum $k \sim T(t_{\rm prod})$ and thereafter redshift as $1/a$, which implies $k(t) \sim T(t)$ for times after production. As a benchmark estimate of the mass threshold $m$ below the KRLP is `warm', we have $m < k(t_{\rm BBN}) \sim T_{\rm BBN} \sim 0.1 - 10 {\rm MeV}$ in order for the particles to be mildly relativistic at BBN.  For 100\% of the DM produced via freeze-in, we expect constraints similar to \cite{Dvorkin:2020xga}. However, the KRLPs are less constrained if only a subcomponent of the dark matter. 

\subsubsection{`Axion-like' Portal}

We now consider an `axion-like' interaction for the KR field, as discussed in Sec.~\ref{sec:Interactions}. Here, the KR interacts with SM  fermions via an axial coupling, given by Eq.~\eqref{eq:4.2}. This interaction is analogous to the axion coupling to the axial current, $\mathcal{L}_{\inter} = \partial_\mu \phi j^{\mu 5}$. This interaction was studied in \cite{Langhoff:2022bij} and shown to lead to an irreducible background for the axion, $\phi$, from freeze-in. We consider again freeze-in via the processes shown in Table~\ref{tab:feynman-diagrams}, whose explicit cross sections can again be found in Appendix~\ref{App:CrossSections}. Fig. \ref{fig:axial-parameters} shows the mass-coupling parameter space for KRLPs freezing in via an axial coupling to fermions, and how these parameters scale for couplings to different fermionic species.

\begin{figure}[ht!]
    \centering
    \includegraphics[width=0.99\linewidth]{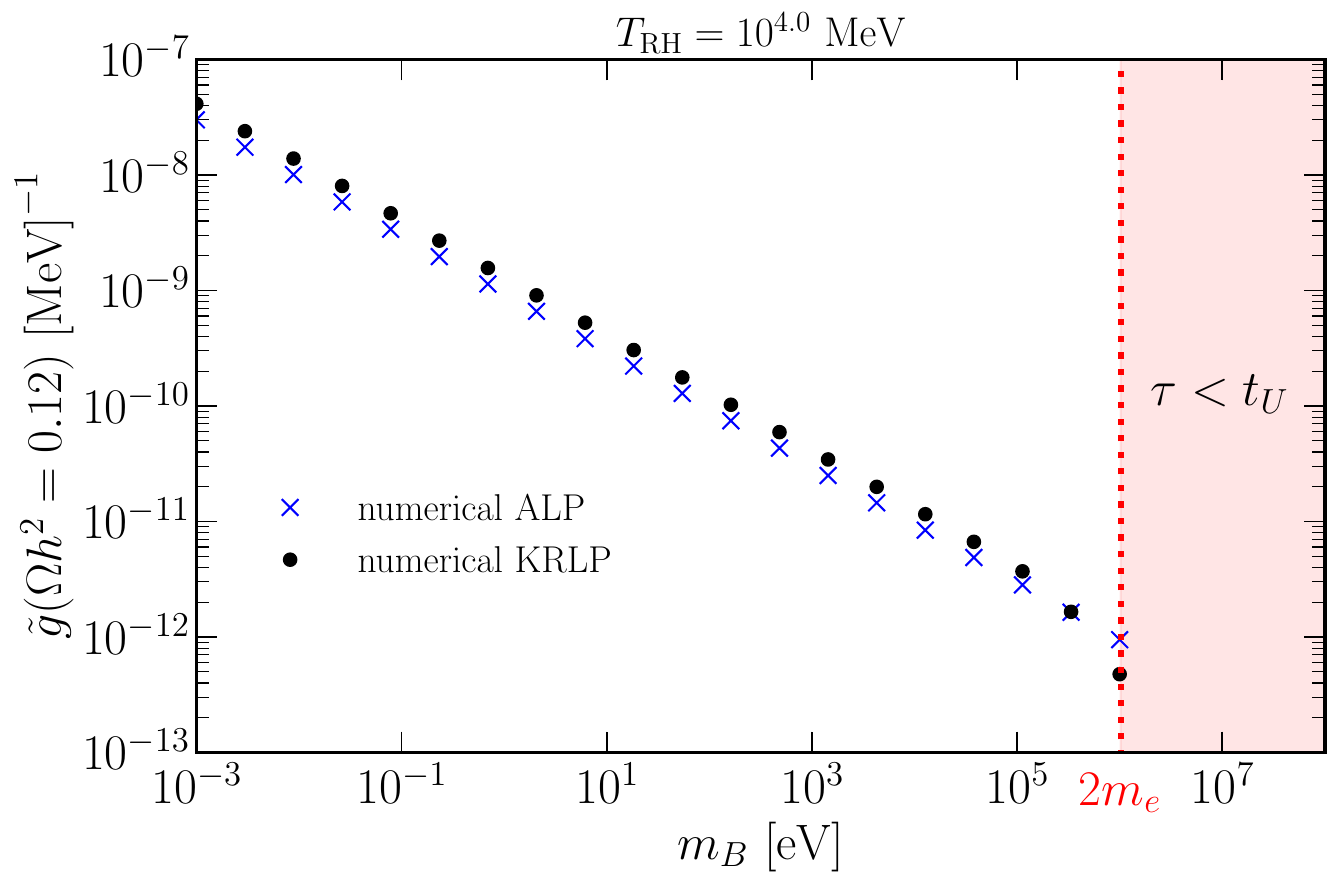}
    \caption{Parameter space for freeze-in via coupling to the electron axial current $j_5^{\mu}$. Here we compare the fractional density of KRLPs (black) to ALPs (blue). The vertical dashed line corresponds to the threshold mass $m=2m_e$, while the shaded red region denotes relics with lifetimes $\tau < 1/H_0$. The values of $\tilde{g}$ in the figure are well within the region where freeze-in is operative.
  }
    \label{fig:axial-parameters}
\end{figure}

 It is interesting to note that production is suppressed at low masses, in contrast with the dark photon-like portal. This happens even though the axial current is not conserved, but instead satisfies $\partial_\mu j^{\mu 5}= 2 i m \bar{\psi} \gamma^5 \psi$ for a massive Dirac fermion $\psi$ \cite{Schwartz:2014sze}, with additional non-conservation from the chiral anomaly. Instead, the relative suppression at small mass follows from the structure of the interaction: given the Lagrangian Eq.~\eqref{eq:4.2}, the interaction vertex is,
 \begin{equation}
     -i\tilde{g}\tilde{V}^{\mu\nu}(k) = -i\tilde{g}\epsilon^{\mu\nu\sigma\rho}\gamma_5\gamma_{\sigma} k_{\rho}
 \end{equation}
where $k_\rho$ is the KRLP momentum. When contracted with an outgoing KRLP, this vanishes due to the simple property,
\begin{equation}
    \epsilon_{\mu \nu \sigma \rho}k^\mu k^\nu =0 .
\end{equation}
This property can be mostly easily understood by rotating to the rest frame of the outgoing KRLP, in which case, in which case the contraction simplifies to $    \epsilon_{\mu \nu 0 0}k^0 k^0= m^2 \epsilon_{\mu \nu 0 0}$ which vanishes due to repeated indices. This guarantees that the momentum-dependent terms in the sum over polarization states (Eq.~\eqref{eq:KRpolsum}) are annihilated when contracted with the vertices.

Thus, from Fig.~\ref{fig:axial-parameters} one can appreciate that the KRLPs in the axion-like interaction freeze-in with the same qualitiative behavior as their axion counterparts as discussed in \cite{Langhoff:2022bij}. The characteristic scaling of the KRLP production, which is independent of $T_\rh$, but dependent on the $m_f$, can be understood in the same way as the dark photon case, discussed above. Explicitly, the arguments made to describe the scaling of the dark photon in Appendix \ref{App:scaling} also apply to the ALP and the axially coupled KRLP shown in Fig.~\ref{fig:axial-parameters}.

\subsubsection{Higgs Portal}

Another production mechanism one could consider for KRLP dark matter is thermal freeze-in via a Higgs portal, as in \cite{Kolb:2017jvz}. The relevant interaction term would be 
\be 
\mathcal{L}_\inter \supset \kappa m_B^2 B_{\mu\nu}B^{\mu\nu}\Phi^\dagger \Phi,
\ee 
where $\kappa$ is a coupling parameter of mass dimension $-2$, and $\Phi$ is the Higgs. This interaction term leads to an effective mass for $B_{\mu\nu}$. As in the above case for GPP, we can immediately see that the freeze-in production via coupling to the Higgs will be identical to the spin-1 counterpart studied in \cite{Kolb:2017jvz}. Thus, once again, we find that one can indeed obtain the correct relic density for the KR to make up the dark matter for a wide range of masses in analogy to a superheavy spin-1.

\section{Gravitational Production of KRLPs \label{sec:other}}

Cosmological gravitational particle production (GPP) is a process by which dark matter is produced non-thermally from quantum fluctuations during the inflationary era due to the rapid expansion of the universe \cite{Parker:1969au, Parker:1971pt, Ford:2021syk,Chung:1998zb,Chung:1998ua}. GPP is an attractive DM production mechanism because it does not require any coupling to the SM, and has been shown to yield the correct relic density for a wide range of particle masses and spins \cite{Chung:1998zb, Kolb:2020fwh, Alexander:2020gmv,Kolb:2021xfn,Kolb:2021nob,Kolb:2023dzp} and in both single field and multifield inflation models \cite{Kolb:2022eyn}. It has previously been shown that GPP is a viable mechanism for production of `completely dark photon' dark matter \cite{Kolb:2020fwh, s1-NM}, characterized by the de Broglie-Proca action with nonminimal couplings to gravity 
\be
S = \int d^4x \sqrt{-g}\left[ -\frac{1}{4} F^{\mu\nu}F_{\mu\nu} + \frac{1}{2}m^2g^{\mu\nu}A_\mu A_\nu - \frac{1}{2}\xi_1 R g^{\mu\nu}A_\mu A_\nu - \frac{1}{2}\xi_2 R^{\mu\nu} A_\nu A_\nu\right],  
\ee
where $R$ is the Ricci scalar, $R^{\mu\nu}$ the Ricci tensor and $\xi_1$ and $\xi_2$ are the coupling constants, analogous to the well-studied nonminimal coupling of scalar fields $\xi \phi^2 R$, which is extensively used in the inflation literature (e.g., \cite{McDonough:2020gmn,Geller:2022nkr,Qin:2023lgo}).

The minimally-coupled KR theory will clearly be dual to the minimally-coupled Proca-de Broglie; however the with the inclusion of the nonminimal couplings the duality is not so obvious. The effect of $\xi_1, \xi_2 \neq 0$ is to lead to effective mass terms, rather than just $m^2$. For the spin-1 DM on an FRW background, these effective masses are \cite{Kolb:2020fwh}
\begin{align}
    m_{\eff, t}^2 &= m^2 - \xi_1 R - \frac{1}{2}\xi_2 R - 3 \xi_2 H^2, \\
    m_{\eff, x}^2 &= m^2 - \xi_1 R - \frac{1}{6}\xi_2 R + \xi_2 H^2.
\end{align}
Now let us consider the analogous KRLP theory. We consider the following action: 
\be 
S = \frac{1}{12}\int d^4x \sqrt{-g}\Big[H_{\mu\nu\rho}H^{\mu\nu\rho} - 3m_B^2B_{\mu\nu}B^{\mu\nu} - \xi_3 R B^{\mu\nu}B_{\mu\nu} - \xi_4 R^{\mu\nu\rho\sigma}B_{\mu\nu}B_{\rho\sigma}\Big], 
\ee 
where $\xi_3$ and $\xi_4$ are new coupling constants. Notice that due to the antisymmetry of $B_{\mu\nu}$ we now cannot have the coupling to the Ricci tensor as in the spin-1 case. We instead consider a coupling to the Riemann tensor. Then, the effective KRLP masses arising from the above action are 
\begin{align}
    m^2_{\eff,t} &= m_B^2 - \frac{2}{3}\xi_3 R -\frac{2}{9}\xi_4 R - \frac{4}{3} \xi_4 H^2  \\
    m^2_{\eff,x} &= m_B^2 - \frac{2}{3}\xi_3 R - \frac{2}{27}\xi_4 R + \frac{4}{3} \xi_4 H^2.
\end{align}
Notice that with a redefinition of the coupling parameters $\xi_3 \rightarrow \frac{3}{2}\xi_1$ and $\xi_4 \rightarrow \frac{9}{4}\xi_2$, the effective masses exactly reduce to those in the massive dark photon scenario. Thus, even in the non-minimally coupled theory, the duality between the GPP of dark photon dark matter and KRLP dark matter does hold, despite having different couplings to gravity.

In this scenario, the duality to the vector dark matter is due to the fact that FRW space is conformally flat. Explicitly, if the Weyl curvature of a spacetime vanishes, then the Riemann tensor can be written in terms of the Ricci tensor and thus the two coupling terms do indeed reduce to the same form. 
 Thus, without needing to perform any calculations for the GPP of KR fields, we can see that the dynamics and relic density of the KR field produced via GPP will be identical to the massive spin-1 field. Thus, the results from \cite{Kolb:2020fwh, s1-NM} for GPP of minimally and non-minimally coupled spin-1 particles, respectively, hold, indicating the the KR field is a viable dark matter candidate for a wide range of masses and coupling parameters. For convenience, we summarize the relevant results below. 

The relevant GPP formulae can be found in \cite{Kolb:2020fwh} and so here we provide only the essentials. We numerically solve the mode equation for the Fourier modes of pseudovector $B_k/a$, given by
\begin{align}
    (\partial_\eta^2  + \omega^2_T) B_k^T &=0,\\
    (\partial_\eta^2  + \omega^2_L)B^L_k &= 0 , 
\end{align}
where $B_k^{T/L}$ are the transverse and longitudinal modes of the KRLP, respectively, and $\omega_{T,L}$ are the associated frequencies for each mode. We then construct the spectrum of particles as \cite{Kolb:2020fwh}
\begin{equation}
    n_k =\frac{k^3}{2\pi^2}|\beta_k|^2, 
\end{equation}
where $\beta_k$ are the Bogoliubov coefficients, and the comoving number density, $na^3$ as 
\be
na^3 = \int \frac{dk}{k}n_k
\ee 
From the comoving mumber density, the relic density can be computed, assuming a reheating temperature and thermal history. We leave these latter analyses to future work, and here perform a preliminary analysis of GPP of the non-minimally coupled KRLP.

\begin{figure}[htb!]
    \includegraphics[width=0.49\textwidth]{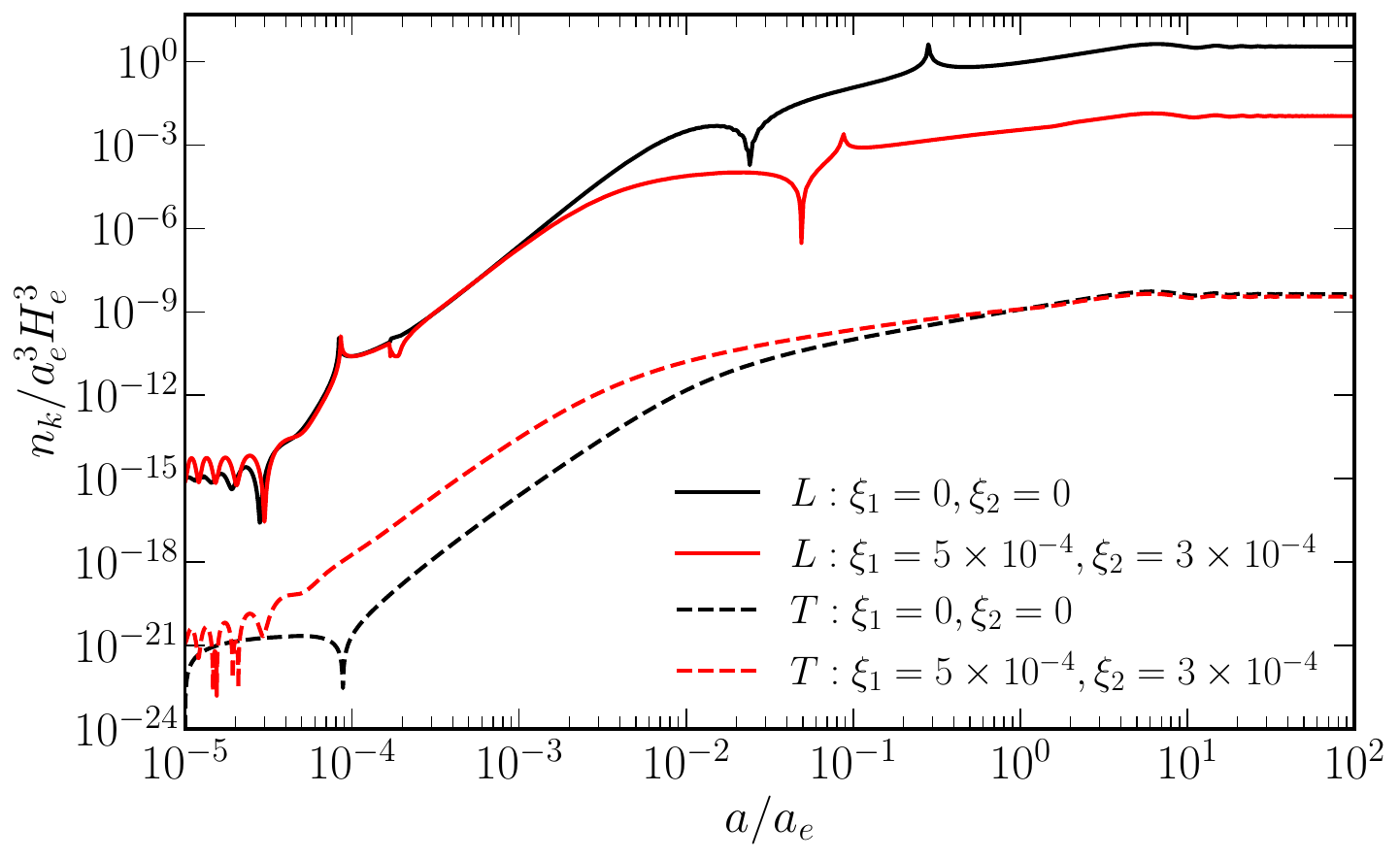}
    \includegraphics[width=0.49\textwidth]{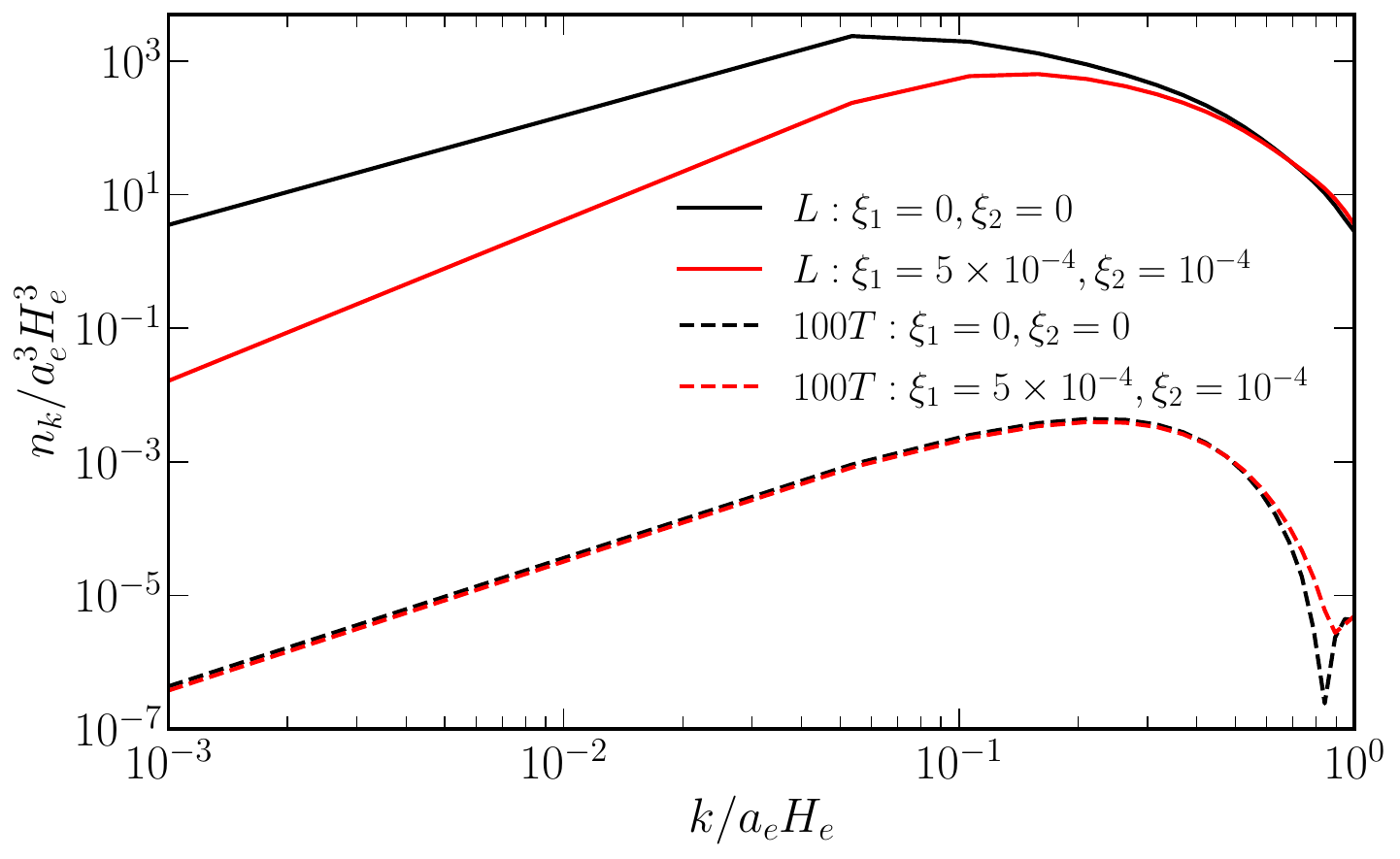}
    \caption{Gravitational particle production for minimally and non-minimally coupled KRLPs for a particular choice of couplings and KRLP mass. ``$L$'' is for the longitudinal mode and ``$T$'' is for the transverse mode.  Note that in the right panel the result for the transverse modes have been multiplied by a factor of 100.}
    \label{fig:GPP}
\end{figure}

In Figure~\ref{fig:GPP} we show the evolution of the KRLP during the inflationary epoch and the resulting number density of produced dark matter, both as a function of the scale factor and as a function of the wavenumber. We show a representative example for $m_B/H_e = 0.1$, where $H_e$ is the Hubble parameter at the end inflation, and $k=0.001$ in the left-hand panel. The black curves show the minimally coupled ($\xi_1 = \xi_2 = 0$) particle number density for both the longitudinal (solid) and transverse (dashed) modes, while the red curves show the same for a particular example of the nonminimallly coupled KRLP, taking $\xi_1 = 5\times 10^{-4}$ and $\xi_2 = 10^{-4}$. Notice that in the right-hand panel, the scaling for the transverse modes has been multiplied by a factor of 100. We see that GPP of KRLPs does indeed lead to significant particle production which can be converted into a present day relic density. Thus, GPP is a viable production mechanism for KRLP dark matter.

\section{FAQ}
\label{sec:FAQ}

In this section we address specific questions regarding the detection prospects for KRLPs, and strategies to distinguish KRLPs from axions or dark photons if detected. 

\subsection{How to experimentally distinguish a Kalb-Ramond from a Dark Photon?}

{\bf Answer:} \emph{Photon-dark photon Oscillations}:
As we have shown above, the KRLP is dual to a massive pseudo-vector field and shares many properties of a dark photon. Thus one may be concerned by the prospect of experimentally distinguishing a KRLP signal from a dark photon signal. The search for dark photons is ongoing at experiments such as Belle-II \cite{Belle-II:2018jsg} 
 and DarkLight \cite{Balewski:2014pxa}, which aim to detect dark photons via $e^+ e^-$ and $e^- p$ collisions, respectively. One again, we emphasize that although the KRLP possesses a duality to a pseudovector field, the KRLP and dark photon are indeed physically inequivalent. Recall that the Kalb-Ramond $B_{\mu \nu}$ is parity-{\it even}, where as the dark photon, $A_{\mu}$, is parity-{\it odd}. Additionally, the dark photon field strength is a rank-two tensor, whereas the Kalb-Ramond field strength is a rank-three tensor.

As discussed in Sec.~\ref{sec:Interactions}, the Kalb-Ramond field is expected to have a magnetic dipole like coupling ${\cal L} \sim B_{\mu \nu}\bar{\psi}\sigma^{\mu\nu}\psi$, in comparison to the vector current coupling of the dark photon, ${\cal L}\sim A_\mu j^\mu$. Both interactions can lead to a displaced vertex, which is the key observable that many of these experiments, such as Belle-II and DarkLight, aim to observe. Assuming no other information other than a binary (non-)detection, this poses a challenge to the identification of the discovered particle as a dark photon or a Kalb-Ramond.

However, a key distinction between the dark photon and KRLP scenarios is that because the dark photon field strength is a rank-two tensor, it can kinetically mix with the electromagnetic field strength as $(F_{\mu})^{\rm DM} (F^{\mu \nu})^{\rm SM}$, whereas for the KRLP, no such kinetic mixing is possible. Thus, for the dark photon, we would expect there to be production via photon-dark photon oscillations, a process which does not occur for KRLPs, and thus if observed, would be a distinguishing feature. Furthermore, due to the characteristic downward slope of the mass-coupling parameter space at low masses for KRLPs, dark photons are able to have a much larger coupling strength than KRLPs. Thus, if one for example detected a $m = 10^{-2}$ eV particle with coupling strength $g = 10^{-14}$ GeV$^{-1}$, the KRLP scenario is automatically ruled out and one can be certain that the detected particle is indeed a dark photon. 

In addition to the potential for individual detection via  dark photon experiments, another distinctive signature of KRLPs which differs from the dark photon scenario would be two seemingly coincident dark photon and axion signatures. If there is a dark photon signal which is not ruled out as a KRLP due to the presence of photon-dark photon oscillations, one could then search for an associated axion signal. In this case, the observations could be explained by as detections of the two KRLP-SM portals rather than needing two dark matter particles to explain the signals.

\subsection{How to experimentally distinguish a KRLP from an axion?}

{\bf Answer}: \emph{The vector-like modes of the Kalb-Ramond, axion coupling to photons}:
Similarly to the dark photon case, let us consider how to experimentally distinguish axions and KRLPs produced via the axion-like portal. Recall that in the massless limit, the Kalb-Ramond field is dual to an axion. Away from the massless limit, the KR field also propagates a vector field. Thus, while a massless KRLP \textit{is} a massless axion, for the massive KRLP the duality does not hold and we obtain the pseudo-vector as discussed above. Recall that in the axion-like coupling, it is actually the Hodge dual of the field strength of the KRLP which couples to the axial current, in contrast with the axion coupling, in which the current simply couples to the derivative of the field. Thus, if an experiment were to observe an axion-like signal and distinguish whether it was coming from a scalar or a pseudo-vector, axions and KRLPs could be differentiated.

Similarly, many experiments are searching for axion signals arising from their characteristic coupling to photons $\mathcal{L} \supset g_{a\gamma\gamma}a F_{\mu\nu}\tilde{F}^{\mu\nu}$. For example, experiments, such as ADMX \cite{ADMX:2009iij} and CAST \cite{Adair:2022rtw} search for axions using a resonant microwave cavity to study the conversion of dark matter axions to microwave photons.  Analogously to the dark photon comparison, KRLPs cannot couple directly to the electromagnetic $F_{\mu\nu}\tilde{F}^{\mu\nu}$ and therefore will not undergo photon conversion in the same way that the axion will. Therefore, if a signal is detected in this way, one can be certain that it has arisen from an axion and not a KRLP. 

On the other hand, the both axions and KLRPs are expected to couple to the axial nuclear moment. The couplings for the axion are to $\partial_\mu\theta$, while the analagous KRLP couplings are to the dual of the field strength, $\tilde{H}_\mu$.  Next generation axion helioscope experiments, such as the International Axion Observatory (IAXO), may be sensitive to axions with nuclear couplings \cite{DiLuzio:2021qct}, and by extension, sensitive to KRLPs. For these experiments, it may be possible to distinguish the axion from a KRLP by focusing on the pseudoscalar vs. pseudo-vector nature of the observed signal. Additionally, as discussed above the dual detection of an axion and dark photon both coupled to electrons could be explained by KRLPs. 

Another avenue is dark matter direct detection experiments. Previous studies have shown that dark matter direct detection experiments which include sensitivity to the angular dependence of nuclear recoils can be used to distinguish the spin of particle dark matter \cite{Catena:2017wzu, Catena:2018uae, Jenks:2022wtj}. In particular, it has been suggested that spin-0 and spin-1 dark matter can be differentiated with this method \cite{Catena:2018uae}, making direct detection a viable avenue with which to search for KRLPS.

\subsection{What other cosmological implications could KRLPs have?}

{\bf Answer}: \emph{Warm and cold dark matter components}:
In addition to the single production mechanisms discussed in Sections \ref{sec:freezein} and \ref{sec:other}, it is possible that each channel could produce a subcomponent of dark matter. This is particularly notable, because freeze-in of KRLPs in the sub-MeV mass range actually yields a warm dark matter component, which can have a wide range of observable consequences as discussed in \cite{Dvorkin:2020xga}. In addition to the warm DM component generated during via freeze-in, a standard cold DM (CDM) could be produced via GPP.  

Here we focused on \emph{stable} KRLPs, which requires $m_B< 2m_f$.  However we note that unstable but long-lived KRLPs would have cosmological effects even if they do not contribute to the present dark matter density.

\section{Discussion and Conclusions \label{sec:Conclusions}}

In this work we have introduced the Kalb-Ramond field and a broader class of Kalb-Ramond-like particles as dark matter candidates. We have discussed the origins of these particles from string theory and derived their interactions with SM particles. Then, we demonstrated that KRLPs can be produced via freeze-in in an analogy to dark photons and axions. We find that there is a distinct difference between freeze-in production of dark photons and KRLPs, arising from the couplings to different SM currents in each case. This leads to a potential overproduction of KRLPs at low masses, and pushing the allowed parameter space to lower couplings. We find for the axion-like freeze-in, Higgs portal freeze-in, and gravitational particle production, that production of KRLPs is qualitatively the same as the axion and vector dark photon scenarios, respectively. 

As we have shown, the KRLP scenario is well motivated and lays the foundation for significant future study. In particular, one may be curious about the observational prospects which we have addressed in the previous section. In addition to direct observations of KRLPs via dark matter experiments, KRLPs may also contribute in other cosmological settings. For example, recall that KRLPs can couple to extended objects. One such possibility is that the KR could interact with string-like defects such as dark matter vortices and leave an observable imprint,  e.g., on the vortices studied \cite{Alexander:2021zhx} and \cite{Alexander:2019puy}. Similarly, if the KRLPs coupled to cosmic strings in the early universe, the oscillation and subsequent decay of the cosmic string loops could induce an observable gravitational wave spectrum. Furthermore, if the KRLP has interactions during inflation, it is possible that these interactions will leave a distinctive signature in the non-Gaussianity of the cosmic microwave background, i.e., a `cosmological collider' signature which could be distinguishable from the usual spin-1 scenario \cite{Chen:2009zp, Arkani-Hamed:2015bza,Lee:2016vti,Alexander:2019vtb}. We also note that the fact that the KRLP has tree-level dipole interactions could have potential implications for the $g-2$ anomalies. Overall, there are many directions forward to further our understanding of KRLPs in cosmology. We leave these studies and others for future work.  

\acknowledgments

The authors thank Stephon Alexander, Keshav Dasgupta, Andrew Frey, Saniya Heeba, Anamaria Hell, Liam McAllister, David McKeen, Hugo Sch\'erer, and Katelin Schutz, for helpful discussions. C.C. acknowledges support from the Canadian Institute for Particle Physics (IPP) via an Early Career Theory Fellowship, and thanks the  Kavli Institute for Cosmological Physics at the University of Chicago for hospitality while a portion of this work was completed.  C.C.\ is supported by a fellowship from the Trottier Space Institute at McGill via an endowment from the Trottier Family Foundation, and by the Arthur B. McDonald Institute via the Canada First Research Excellence Fund (CFREF). The work of L.J.\ is supported by the Kavli Institute for Cosmological Physics at the University of Chicago via an endowment from the Kavli Foundation and its founder Fred Kavli. The work of E.W.K. was supported in part by the US Department of Energy contract DE-FG02-13ER41958 and the Kavli Institute for Cosmological Physics at the University of Chicago.  EM is supported in part by a Discovery Grant from the Natural Sciences and Engineering Research Council of Canada.

\begin{appendix}

\section{Quantization of the Kalb-Ramond Field}
\label{app:KR}

Here we explicate the technical details of the quantization of the KR field in Minkowski space necessary to calculate the freeze-in cross sections. First, to perform a mode expansion of $B_{\mu \nu}$, it is useful to expand into a basis of solutions of the constraint equation $\partial_\mu B^{\mu \nu}=0$ (which itself follows from the equation of motion).  We write $B_{\mu \nu}$ as
\begin{equation}
    B_{\mu\nu}(x)=\int\frac{d^3k}{4\sqrt{2\pi E}}\varepsilon^ {(\alpha)}_{\mu \nu} \big(a^{\dagger(\alpha)}(k)e^{ik\cdot x}+h.c. \big)
\end{equation}
where the polarization tensors $\varepsilon^ {(\alpha)}_{\mu \nu}$, with $\alpha=1, 2, 3$ (or equivalently $\pm1$ and $0$), are defined by the 4d transverse condition, $k_{\mu}\varepsilon^{\mu \nu} = 0$, which follows from the constraint $\partial_\mu B^{\mu \nu}=0$. The polarization tensors also inherit antisymmetry and tracelessness from $B_{\mu \nu}$. To obtain explicit expressions for the polarization tensors, we expand the Proca field, $B_k$ in spin-1 polarization vectors $\varepsilon_{k}$, and from this construct the spatial part of the polarization tensors $\varepsilon_{\mu\nu}$ for the KR field as
\begin{equation}
\varepsilon_{ij} ^{(\lambda)} = \epsilon_{ijk} \varepsilon^{(\lambda)k}    
\end{equation}
where $\lambda=0$ denotes the polarization tensor of the longitudinal mode and  $\lambda=\pm 1$ denotes the polarization state of the transverse modes, and $\varepsilon^{(\lambda) k}$ are the usual Proca polarization vectors. Then, taking the usual expressions for $\varepsilon_k^{(\lambda)}$ to be $\varepsilon_k^{\pm 1} = (1, \pm i, 0)$ and  $\varepsilon_k^0 = (0,0,1)$ and $k_\mu = (E, 0, 0, k)$ for a massive vector field, we obtain 

\begin{align}
 \varepsilon_{\mu\nu}^{(0)} = \frac{1}{\sqrt{2}}
\begin{pmatrix}
    0 & 0 & 0 & 0\\
    0 & 0 & -1 & 0\\
    0 & 1 & 0 & 0 \\
    0 & 0 & 0 & 0
\end{pmatrix} \,\, , \,\, & 
\varepsilon_{\mu\nu}^{\pm 1} = 
\frac{1}{2\sqrt{1-k^2/E^2}}\begin{pmatrix}
0 & \pm i k/E & -k/E & 0 \\\mp i k/E & 0 & 0 & \mp i   \\ k/E & 0 & 0 & 1 \\ 0 & \pm i  & -1 & 0
\end{pmatrix}. 
\label{eq:polTens}
\end{align}
From the polarization tensors in Eq.~\eqref{eq:polTens}, one can find the polarization sum necessary to calculate cross sections, which is given by: 

\begin{equation}
\label{eq:KRpolsum}
\sum_{\lambda}\varepsilon_{\mu\nu}^{\dagger(\lambda)}\varepsilon_{\rho\sigma}^{(\lambda)} = -\frac{1}{2}\left(g^{\mu \sigma} g^{\rho \nu}- g^{\mu \rho} g^{\nu \sigma} \right)
 -\frac{1}{2m^2}\left(g^{\mu \rho} k^\nu k^\sigma-g^{\nu \rho} k^\mu k^\sigma-g^{\mu \sigma} k^\nu k^\rho + g^{ \nu \sigma} k^\mu k^\rho \right).
\end{equation}
This can be derived by any of three ways: (1) directly from the polarization tensors, or (2) directly from the action \cite{Pilling:2002dz}, or instead (3) proposed as an ansatz that satisfies all required symmetries (transverse, antisymmetry in indices) and moreover is consistent with the orthogonality of the polarization tensors, namely that $P^{\mu \nu \rho \sigma} \equiv \sum_{\lambda}{\varepsilon^{\mu\nu}}^{\dagger(\lambda)}{\varepsilon^{\rho\sigma}}^{(\lambda)} $ satisfies $\varepsilon^{\dagger \rho \sigma +} P_{\mu \nu \rho \sigma} = \varepsilon ^{\dagger +} _{\mu \nu}$, and similarly for the $\lambda=0$ and $\lambda=-$ polarization states .

With the above expressions, we now have all of the necessary tools in order to calculate cross sections for freeze-in processes.  

\section{Cross Section Calculations}
\label{App:CrossSections}
In this appendix, we discuss explicit details of the cross section computations for each of the processes contributing to the KRLP freeze-in relic density, for both the dark photon-like interaction and the axion-like interaction. In the calculations that follow, we make use of the Mathematica package FeynCalc \cite{Mertig:1990an} to simplify Dirac trace expressions.

\subsection{Dark Photon-Like Portal}
We first consider the dark photon-like portal for KRLP freeze-in, described by the interaction Lagrangian Eq.~\eqref{eq:3.1}. This interaction gives the vertex:
\begin{equation}
    -igV^{\mu\nu}=-g\sigma^{\mu\nu},
\end{equation}
where we will take $g\lesssim e\ll1$. Consider freeze-in through the Fermion $f$. The contributing processes are shown in Table \ref{tab:feynman-diagrams}. The leading order process $\bar{f}f\rightarrow B$ (sometimes called ``coalescence" or ``inverse decay") is kinematically forbidden for $m_B<2m_f$, and as such we must consider $\mathcal{O}(eg)$ diagrams for masses smaller than $2m_f$.  One is a Compton-like scattering process $f \gamma \rightarrow f B$, and the other is the annihilation $\bar{f}f\rightarrow \gamma B$. 

Let us first compute the coalescence cross section. The amplitude for coalescence is given by 
\begin{equation}
    -i\mathcal{M}(\bar{f}f\rightarrow B) = \bar{v}(s';p_2)(-igV^{\mu\nu})u(s;p_1)\varepsilon^{\dagger(\lambda)}_{\mu\nu}. 
\end{equation}
In the center of mass frame, $p_1=(\sqrt{s}/2,\ \vec{p})$, $p_2=(\sqrt{s}/2,\ -\vec{p})$, and $k=(m_B,\ 0)$. Also, $k^2=m_B^2$, $p_1^2=p_2^2=m_f^2$, $(p_1\cdot p_2) = s/2-m_f^2$, and $(k\cdot p_1)=(k\cdot p_2) = m_{B}\sqrt{s}/2 $. Summing over incoming and outgoing spin states, denoted by an overbar $\bar{( \ )}$, and using these kinematic identities, we arrive at the quantity :
\begin{equation}
    \overline{|\mathcal{M}|}^2 = \sum_{s,s'}\sum_{\lambda} \mathcal{M}^{\dagger\mu\nu}\mathcal{M}^{\rho\sigma}\varepsilon^{\dagger(\lambda)}_{\mu\nu}\varepsilon^{(\lambda)}_{\rho\sigma} = 4g^2\left(s-4m_f^2 \right),
\end{equation}
where $s$ is the center of mass Mandelstam variable. Note that here we have used the identity in Eq. \ref{eq:KRpolsum} in order to simplify the contraction with external KR polarizations. We may now compute the cross section $\bar{\sigma}(s)$ from this amplitude. For a $2\rightarrow 1$ process, we have that \cite{Gondolo:1990dk}
\begin{equation}
    \bar{\sigma}(s)\frac{s}{2}\sqrt{1-\frac{4m_f^2}{s}} = \frac{\pi}{2}\overline{|\mathcal{M}|}^2\delta(s-m_B^2),
\end{equation}
from which the coalescence cross section may be read off as:
\begin{equation}
    \bar{\sigma}(\bar{f}f\rightarrow B)(s) = 4\pi g^2\beta_1\delta(s-m_{B}^2), 
\end{equation}
where we have defined the quantity $\beta_1 = \sqrt{1-4m_f^2/s}$.

Next, consider $\mathcal{O}(eg)$ processes. At this order, each process has two contributing diagrams. The annihilation cross section $\bar{\sigma}(\bar{f}f\rightarrow\gamma B)$ has a t-channel and u-channel contributions, where the t-channel amplitude is given by
\be
    -i\mathcal{M}_t(\bar{f}f\rightarrow\gamma B) = \bar{v}(p_2)\left(-i g V^{\alpha\beta}\right) \frac{i\left(\slashed{p}_1-\slashed{p}_3+m_f\right)}{t-m_f^2}\left(-ie\gamma^{\mu}\right)u(p_1)\varepsilon^{\dagger(\lambda)}_{\alpha\beta}\varepsilon^{(\lambda')}_{\mu},
\ee
where we've defined the system such that $p_1^2=p_2^2=m_f^2$, $p_3^2=0$, and $p_4^2=m_B^2$. The u-channel amplitude may be found by exchanging the order of the vertices, sending $p_3\rightarrow p_4$, and $t\rightarrow u$. The overall matrix element squared will involve now four terms:
\begin{equation}
\begin{split}
    \overline{|\mathcal{M}|}^2 = \sum_{s,s'}\sum_{\lambda,\lambda'}\left( \mathcal{M}^{\dagger}_t\mathcal{M}_t+\mathcal{M}^{\dagger}_u\mathcal{M}_u +\mathcal{M}^{\dagger}_t\mathcal{M}_u + \mathcal{M}^{\dagger}_u\mathcal{M}_t \right).
\end{split}
\end{equation}
After making use of Dirac algebra and kinematic relations, we arrive at:
\begin{equation}
    \begin{split}
        \overline{|\mathcal{M}|}^2 & =\frac{8 e^2 g^2}{m_B^2 (m_f^2-t)^2 (m_f^2-u)^2} \Bigg[-4 m_f^6 \left(4 m_B^2 (t+u)+3 m_B^4-4 \left(t^2+3 t u+u^2\right)\right) \\ 
        &+m_f^4
   \left(m_B^2 \left(5 t^2-2 t u+5 u^2\right)+10 m_B^4 (t+u)+2 m_B^6-4 (t+u) \left(t^2+8 t
   u+u^2\right)\right) \\
   & +m_f^2 \left(-2 m_B^6 (t+u)-4 t u m_B^4-m_B^2 (t-u)^2 (t+u)+8 t u \left(t^2+3 t
   u+u^2\right)\right) \\
   & +20 m_f^8 \left(m_B^2-t-u\right)+t u \big(-2 m_B^4 (t+u)+m_B^2 (t+u)^2+2 m_B^6  \\
   &  -4 tu (t+u)\big)+8 m_f^{10}\Bigg]
    \end{split}
\end{equation}
\begin{equation}
    \begin{split}
        \bar{\sigma}(f\bar{f}\rightarrow\gamma B)(s) = \frac{\alpha  g^2 }{m_B^2
   s^2 \left(s-m_B^2\right) \beta_1^2}\Big[-2 \beta_1  \left(m_B^2 s \left(3 s-4 m_f^2\right)+m_B^6-2 m_B^4 s-s^3\right) &\\
    +m_B^2  \left(-4 m_f^2 \left(m_B^2+2 s\right)+16
   m_f^4+m_B^4+s^2\right)\log \left(\frac{1+\beta_1}{1-\beta_1} \right)\Big].&
    \end{split}
\end{equation}
In the limit $s\gg (m_f^2,m_B^2)$, the cross section $\bar{\sigma}(f\bar{f}\rightarrow\gamma B)(s) \to \alpha g^2/m_B^2$.

As for the Compton-like process ($f\gamma\rightarrow f B$), the diagrams are related to Fermion annihilation by a crossing symmetry. This allows us to quickly compute $\overline{|\mathcal{M}|}^2(f\gamma\rightarrow f B)$ by letting $t \rightarrow -s$, and multiplying by an overall factor of $-1$. Integrating this amplitude in the same fashion as before, we find
\begin{equation}
    \begin{split}
         \bar{\sigma}(f\gamma & \rightarrow f B)(s) = \frac{\alpha  g^2 }{s m_B^2 \left(s-m_f^2\right)^3}\Bigg[\beta _2 \Big(m_f^4 \left(7 s m_B^2-m_B^4+24 s^2\right)\\
         &+s m_f^2 \left(-37 s m_B^2+2 m_B^4-16 s^2\right) 
         -m_f^6 \left(3 m_B^2+16 s\right)  \\ & +s^2
   \left(s m_B^2+7 m_B^4+4 s^2\right)+4 m_f^8\Big) \\
    &+2 s m_B^2 \big(m_f^2 \left(6 s-10 m_B^2\right)-2 s m_B^2+2 m_B^4+9 m_f^4+s^2\big) \\
    & \times \log \left(\frac{m_f^2-m_B^2+s(1+\beta_2)}{m_f^2-m_B^2+s(1-\beta_2)}\right)\Bigg]
    \end{split}
\end{equation}
where we have defined the quantity $s^2\beta_2^2 = (s-(m_f+m_B)^2)(s-(m_f-m_B)^2)$. Again, in the limit $s\gg (m_f^2,m_B^2)$, the cross section $\bar{\sigma}(f\gamma\rightarrow f B)(s) \to \alpha g^2/m_B^2$.

\subsection{Mass-Dependence From the Polarization Sum}
Here we show explicitly that there is a mass-dependent term that survives in the calculation of $|\mathcal{M}|^2$ for the production of KRLPs through a magnetic dipole interaction. This is in contrast to the dark photon, and results in an enhancement of production of KRLPs for lower masses (pictured in Fig. \ref{fig:dipole-parameters}). For concreteness, let us compare the Compton-like photon conversion process $f \gamma \rightarrow f X$, where $X$ will stand for either the dark photon or KRLP.
We may factor a the dark photon matrix element as
\begin{equation}
    \overline{|\mathcal{M}|}^2=\mathcal{M}^{\dagger \mu\nu}\mathcal{M}^{ \rho\sigma} \left(\sum_{\lambda}\varepsilon^{\dagger(\lambda)}_{\mu}(p_2)\varepsilon^{(\lambda)}_{\nu}(p_2)\right)\left(\sum_{\lambda'}\varepsilon^{\dagger(\lambda')}_{\rho}(p_4)\varepsilon^{(\lambda')}_{\sigma}(p_4)\right).
\end{equation}
We are only concerned here with the momentum-dependent piece of the polarization sum, defined
\begin{equation}
    K_{\rho\sigma}(k) = \frac{k_{\rho}k_{\sigma}}{m_B^2}.
\end{equation}
Replacing this definition in the second set of parentheses, we have:
\begin{equation}
\begin{split}
    \overline{|\mathcal{M}|}^2 & \rightarrow \mathcal{M}^{\dagger \mu\nu}\mathcal{M}^{ \rho\sigma} \left(-g_{\mu\nu}\right)K_{\rho\sigma}(p_4) =\frac{4 e^2 g^2 }{\left(s-m_f^2\right)^
   2}\Big[(-2 m_f^4 \left(5
   m_B^2-s-t+u\right)\\
   &+m_f^2 \left(-5 m_B^4+m_B^2 (8 s+6
   t+4 u)-3 s^2-4 s t+2 s u-t^2+u^2\right)\\
   & +s
   \left(m_B^4-2 m_B^2 (s+t)+s^2+2 s
   t+t^2-u^2\right)\Big]=0.
\end{split}
\end{equation}
In the last line, we have used that $s=\sum m_i^2-t-u$. Observe that this quantity vanishes, even though the dark photon itself may have a mass $k^2=m^2$. This is a consequence of the vector current $j^{\mu}$ being conserved.

Now turning to Compton-like production of a KRLP, we may factor the matrix element as:
\begin{equation}
    \overline{|\mathcal{M}|}^2=\mathcal{M}^{\dagger \mu\alpha\beta}\mathcal{M}^{ \nu\gamma\delta} \left(\sum_{\lambda}\varepsilon^{\dagger(\lambda)}_{\mu}\varepsilon^{(\lambda)}_{\nu}\right)\left(\sum_{\lambda'}\varepsilon^{\dagger(\lambda')}_{\alpha\beta}\varepsilon^{(\lambda')}_{\gamma\delta}\right).
\end{equation}
In this example, we replace the polarization sum with its momentum-dependent piece defined:
\begin{equation}
    K_{\alpha\beta\gamma\delta}(k) = -\frac{1}{2m^2}\left(g^{\mu \rho} k^\nu k^\sigma-g^{\nu \rho} k^\mu k^\sigma-g^{\mu \sigma} k^\nu k^\rho + g^{ \nu \sigma} k^\mu k^\rho \right).
\end{equation}
Replacing this above, we have:
\begin{equation}
    \begin{split}
    \overline{|\mathcal{M}|}^2  \rightarrow &\mathcal{M}^{\dagger \mu\alpha\beta}\mathcal{M}^{ \nu\gamma\delta} \left(-g_{\mu\nu}\right)K_{\alpha\beta\gamma\delta}(p_4) \\ =& 
        \frac{8 e^2 g^2}{m_B^2
   \left(t-m_f^2\right)^2
   \left(u-m_f^2\right)^2} \Big[-4 m_f^8 \left(7 m_B^2+5
   (t+u)\right)+4 m_f^6 (3 m_B^4+2 m_B^2 (t+u)\\
    &+ 4 \left(t^2+3 t u+u^2\right))+m_f^4 (2
   m_B^6-14 m_B^4 (t+u)+m_B^2 \left(5 t^2+46 t u+5
   u^2\right)\\
   & - 4 \left(t^3+9 t^2 u+9 t
   u^2+u^3\right))-m_f^2 (2 m_B^6
   (t+u)-20 m_B^4 t u \\ 
   &+m_B^2 \left(t^3+23 t^2 u+23 t
   u^2+u^3\right) -8 t u \left(t^2+3 t
   u+u^2\right))+8 m_f^{10} \\
   & +t u \left(2
   m_B^6-2 m_B^4 (t+u)+m_B^2 (t+u)^2-4 t u
   (t+u)\right)\Big].
    \end{split}
\end{equation}
This term is not only finite, unlike the case of the dark photon, but has a $1/m^2$ scaling. The survival of this mass-dependent term is what leads to the enhancement of production at low masses.

\subsection{Axion-Like Portal}
The calculation of cross sections for the axial interaction is identical except for the replacement of the new vertex $gV^{\mu\nu}\rightarrow \tilde{g}\tilde{V}^{\mu\nu}$.
Then, one finds a simple expression for the three-point coalescence process: 
\be 
\overline{|\mathcal{M}|}^2 = 2 \tilde{g}^2 m_B^2 \left(s-4 m_e^2\right),
\ee 
such that
\begin{equation}
    \bar{\sigma}(\bar{f}f\rightarrow B)(s) = 2\pi \tilde{g}^2m_B^2\beta_1\delta(s-m_{B}^2).
\end{equation}
For fermion annihilation, we have now:
\begin{equation}
    \begin{split}
         \bar{\sigma}(f\bar{f}\rightarrow\gamma B)(s) =& \frac{2\alpha  \tilde{g}^2}{s^2 \beta_1^2(s-m_B^2)}\Bigg\{\frac{ m_B^2 (s-4 m_f^2)}{s (s-2 m_f^2)} \Big[2 m_f^2 \left(-2 s m_B^2+m_B^4+3 s^2\right)\\
       &  +s \left(-s  m_B^2+m_B^4+s^2\right)\Big] + \Big[m_B^2 \left(m_B^4+s^2\right)-2 m_f^2 \left(6 s m_B^2+m_B^4-s^2\right)  \\
       & +16 m_B^2 m_f^4\Big] \log \left[ \frac{1+\beta_1}{1-\beta_1}\right]\Bigg\} \, ,
       \end{split}
\end{equation}
and for Compton-like scattering we have

\begin{align}
    \bar{\sigma}(f\gamma\rightarrow\gamma B)(s)= & \frac{\alpha  \tilde{g}^2}{(s-m_f^2)^3} \Bigg\{-\frac{\beta_2}{2 s} \Bigg(m_f^4 \left(-3 s
   m_B^2+m_B^4-14 s^2\right) \nonumber \\&
   +s m_f^2 \left(35 s m_B^2-2 m_B^4+6
   s^2\right)\nonumber\\&+m_f^6 \left(m_B^2+10 s\right)-s^2 m_B^2 \left(7
   m_B^2+s\right)-2 m_f^8\Bigg) \nonumber\\&
    +\Bigg(2 m_f^2 \left(3 s
   m_B^2-5 m_B^4+s^2\right)+m_f^4 \left(9 m_B^2-4 s\right) \nonumber\\& +m_B^2
   \left(-2 s m_B^2+2 m_B^4+s^2\right)+2 m_f^6\Bigg) \log\left[ \frac{m_f^2-m_B^2+s(1+\beta _2)}{m_f^2-m_B^2+s(1-\beta _2)}\right]\Bigg\}.
\end{align}
In the limit where $m_f>m_B$ and $s\gg m_f^2$, the cross sections scale as $\bar{\sigma}\propto \tilde{g}^2 m_f^2/s$.

\section{Relic Density Scaling}
\label{App:scaling}
In this Appendix, we discuss in further detail the characteristic scaling with $T_\rh$ for KRLPs. As we have discussed above, the KRLP cross sections for the dark photon-like $2 \rightarrow 2$ processes scale as $\bar{\sigma} \propto  g^2/m_B^2$ in the large-$s$ limit, while the dark photon, axion, and KRLP axion-like interaction cross sections scale as $\bar{\sigma} \propto g^2 m_f^2/s$ in the large-$s$ limit. To illustrate how this leads to the $m_f$ and $T_\rh$ dependence discussed in Section~\ref{sec:freezein}, consider the specific case of Compton scattering. For the KRLP dipole interaction, the Boltzmann equation in the large-s limit becomes 
\be 
\dot{n}_B + 3 H n_B \propto \frac{T g^2}{m_B^2}\int_{m_f^2}^\infty ds \, K_1\left(\frac{\sqrt{s}}{T}\right)\frac{(s - m_f^2)^2}{\sqrt{s}}. 
\ee 
Performing the integral, we can write the RHS as 
\be 
\bar{n}^2 \langle \sigma v\rangle \propto g^2 \frac{m_f^2 m_B^2}{x^2}K_2\left(\frac{m_f}{m_B}x\right), 
\ee 
where we have again defined $x = m_B/T$. The abundance is then given by 
\be 
Y_B \propto g^2 \frac{m_f^2}{m_B^2}\frac{M_{\pl}}{m_B}\int_{m_B/T_\rh}^\infty K_2\left(\frac{m_f}{m_B}x\right) dx.
\ee 
For small arguments of the Bessel function, $K_2 \approx m_B^2/m_f^2x^2$, and for large arguments it is exponentially suppressed, which yields 
\be 
Y_B \propto g^2 \frac{M_{\pl}}{m_B}\frac{T_\rh}{m_B}.
\ee 
To find the characteristic scaling, consider the critical density $\Omega = m_B Y_B = 1$ to obtain 
\be 
g \propto M_{\pl}^{-1/2}\left(\frac{m_B}{T_\rh}\right)^{1/2}.
\label{eq:scalingKR}
\ee 
Now consider the analogous dark photon scenario in which the cross section $\bar{\sigma}\propto 1/s$, again for Compton scattering. In this case we have 
\be 
\bar{n}^2\langle \sigma v\rangle \propto g^2 T \int_{m_f^2}^\infty ds \, s^{1/2} K_1\left(\frac{\sqrt{s}}{T}\right) = g^2\frac{m_B^2m_f^2}{x^2}K_2\left(\frac{m_f}{T}\right)
\ee 
to obtain 
\be 
Y \propto g^2 \frac{M_{\pl}}{m_B}\frac{m_f^2}{m_B^2}\int_{m_B/T_\rh}^\infty x^2 K_2\left(\frac{m_f^2}{m_B^2}x\right).
\ee 
For large values of $m_f^2x/m_B^2$, $K_2$ is exponentially small such that we can take the upper limit on the integral to be simply $m_B/m_f$. We then arrive at 
\be 
Y \propto g^2 \frac{M_{\pl}}{m_f}.
\ee 
To again obtain the critical value for $\Omega_B$, we find 
\be 
g \propto M_{\pl}^{-1/2}\left(\frac{m_f}{m_B}\right)^{1/2}. 
\label{eq:scalingDP}
\ee 
The overall scaling due to annihilation will follow in a similar way, as does the scaling of the axial processes. Similar arguments can be made to understand the differences between our KRLP dipole interaction and the dipole coupling to the dark photon field strength in \cite{Krnjaic:2022wor}, which finds $g\propto (m_{A'}T_\rh)^{-1/2}$. Thus, from the two expressions Eqs.~\eqref{eq:scalingKR} and~\eqref{eq:scalingDP}, we can analytically understand the behavior of the dark photon and KRLP parameter space in Figures~\ref{fig:dipole-parameters} and~\ref{fig:axial-parameters}. 

\end{appendix}

 \bibliographystyle{JHEP} 
 \bibliography{ref}

\end{document}